\documentclass[twocolumn,journal]{IEEEtran}
\usepackage[T1]{fontenc}
\usepackage[latin9]{inputenc}
\usepackage{array}
\usepackage{amsmath}
\usepackage{amsthm}
\usepackage{amssymb}
\usepackage{graphicx}
\usepackage{esint}
\usepackage[unicode=true,
 bookmarks=true,bookmarksnumbered=true,bookmarksopen=true,bookmarksopenlevel=1,
 breaklinks=false,pdfborder={0 0 0},pdfborderstyle={},backref=false,colorlinks=false]
 {hyperref}
\hypersetup{pdftitle={Your Title},
 pdfauthor={Your Name},
 pdfpagelayout=OneColumn, pdfnewwindow=true, pdfstartview=XYZ, plainpages=false}

\makeatletter

\providecommand{\tabularnewline}{\\}

\theoremstyle{plain}
\newtheorem{thm}{\protect\theoremname}
\theoremstyle{plain}
\newtheorem{prop}[thm]{\protect\propositionname}

\usepackage[caption=false,font=footnotesize]{subfig}
\usepackage{acronym}
\AtBeginDocument{\acrodef{AO}{alternating optimization}
\acrodef{ASP}{antenna separation product}
\acrodef{AWGN}{additive white Gaussian noise}
\acrodef{BEP}{bit error probability}
\acrodef{BER}{bit error rate}
\acrodef{BF-MIMO}[BF\mbox{-}MIMO]{beamforming MIMO}
\acrodef{BF}{beamforming}
\acrodef{bpcu}{bits per channel use}
\acrodef{CP}{cyclic prefix}
\acrodef{CSI}{channel state information}
\acrodef{CSIR}{channel state information at receiver}
\acrodef{SSK}{space shift keying}
\acrodef{CSIT}{channel state information at the transmitter}
\acrodef{DCMC}{discrete\mbox{-}input continuous\mbox{-}output memoryless channel}
\acrodef{DFT}{discrete Fourier transform}
\acrodef{DL-TR-GSM}{dual-layered transmit-receive \acl{GSM}}
\acrodef{DLT}{dual-layered transmission}
\acrodef{EGC}{equal gain combining}
\acrodef{EM}{electromagnetic}
\acrodef{FSPL}{free space path loss}
\acrodef{FFT}{fast Fourier transform}
\acrodef{FDE}{frequency domain equalization}
\acrodef{GRSM}{generalized \acl{RSM}}
\acrodef{GSM}{generalized \acl{SM}}
\acrodef{IFFT}{invserse fast Fourier transform}
\acrodef{ICI}{inter-channel interference}
\acrodef{iid}[i.i.d.]{independent and identically distributed}
\acrodef{IQ}{in\mbox{-}phase and quadrature}
\acrodef{ISI}{intersymbol interference}
\acrodef{ISI-free}[ISI\mbox{-}free]{intersymbol interference free}
\acrodef{LIS}{large intelligent surface}
\acrodef{LOS}{line\mbox{-}of\mbox{-}sight}
\acrodef{mmWave}{millimeter-wave}
\acrodef{MIMO}{multiple\mbox{-}input multiple\mbox{-}output}
\acrodef{MISO}{multiple\mbox{-}input single\mbox{-}output}
\acrodef{ML}{maximum likelihood}
\acrodef{MRC}{maximal ratio combining}
\acrodef{MMSE}{minimum mean square error}
\acrodef{MU-TR-GSM}{multiuser transmit-receive  \acl{GSM} }
\acrodef{NCSIT}{no channel state information at TX}
\acrodef{NLOS}{non\mbox{-}\acs{LOS}} 
\acrodef{NOMA}{non-orthogonal multiple access}
\acrodef{OFDM}{orthogonal frequency division multiplexing}
\acrodef{OFDMA}{orthogonal frequency division multiple access}
\acrodef{PA}{power amplifier}
\acrodef{PAE}{power added efficiency}
\acrodef{PAPR}{peak\mbox{-}to\mbox{-}average power ratio}
\acrodef{PDF}{probability density function}
\acrodef{PEP}{pairwise error probability}
\acrodefplural{PEP}{pairwise error probabilities}
\acrodef{PGM}{projected gradient method}
\acrodef{APGM}{accelerated projected gradient method}
\acrodef{PMP}{probability mass function}
\acrodef{PSM}{precoding-aided spatial modulation}
\acrodef{QSM}{quadrature spatial modulation}
\acrodef{RC}{reorganization computation}
\acrodef{RIS}{reconfigurable intelligent surface}
\acrodef{RSM}{receive spatial modulation}
\acrodef{RX}{receiver}
\acrodef{SEP}{symbol error probability}
\acrodef{SER}{symbol error rate}
\acrodef{SINR}{signal-to-interference-plus-noise ratio}
\acrodef{SISO}{single-input single-output}
\acrodef{SM}{spatial modulation}
\acrodef{SMX-MIMO}[SMX\mbox{-}MIMO]{spatial multiplexing MIMO}
\acrodef{SMX}{spatial multiplexing}
\acrodef{SNR}{signal-to-noise ratio}
\acrodef{SC}{single carrier}
\acrodef{SVD}{singular value decomposition}
\acrodef{SPST}{single pole single-throw}
\acrodef{SU}{secondary user}
\acrodef{TDE}{time domain equalization}
\acrodef{TX}{transmitter}
\acrodef{ULA}{uniform linear array}
\acrodef{URA}{uniform rectangular array}
\acrodef{VGA}{variable gain amplifier}
\acrodef{ZF}{zero-forcing}
\acrodef{ZMCG}{zero-mean complex Gaussian}
\acrodef{SL-MIMO}{sparse layered MIMO}
\acrodef{MP}{message-passing}
\acrodef{MGF}{moment generating function}
\acrodef{AWEP}{average word error probability}
\acrodef{LDS}{low density signature}
\acrodef{SCMA}{sparse code multiple access}
\acrodef{RE}{resource element}
\acrodef{SL}{sparse layering}
\acrodef{SMX}{spatial multiplexing}
\acrodef{VBLAST}{vertical Bell Labs layered space-time}
\acrodef{STBC}{space-time block code}
\acrodef{SMX}{spatial multiplexing}
\acrodef{MSTBC}{multi-dimensional STBC}}
\usepackage{cite}

\makeatother

\providecommand{\propositionname}{Proposition}
\providecommand{\theoremname}{Theorem}

\begin{document}
\title{Sparse Layered MIMO with Iterative Detection}
\author{Mohamad H. Dinan, Nemanja Stefan Perovi\'c,~\IEEEmembership{Member,~IEEE,}
and~Mark F. Flanagan,~\IEEEmembership{Senior~Member,~IEEE}\thanks{This work was funded by the Irish Research Council (IRC) under the
Consolidator Laureate Award Programme (grant number IRCLA/2017/209).}\thanks{Mohamad H. Dinan, Nemanja Stefan Perovi\'c, and Mark F. Flanagan
are with School of Electrical and Electronic Engineering, University
College Dublin, Belfield, Dublin 4, Ireland (Email: mohamad.hejazidinan@ucdconnect.ie,
nemanja.stefan.perovic@ucd.ie, mark.flanagan@ieee.org).}}
\maketitle
\begin{abstract}
In this paper, we propose a novel transmission scheme, called \emph{sparse layered MIMO} (SL-MIMO)
, that combines non-orthogonal transmission and \ac{SVD} precoding.
Non-orthogonality in SL-MIMO allows re-using of the eigen-channels
which improves the spectral efficiency and error rate performance
of the system through enhancing the coding gain and diversity gain.
We also present a low-complexity \ac{MP} detector for the proposed
SL-MIMO system which performs quite close to \ac{ML}. The joint
\ac{MGF} of the \emph{ordered} eigenvalues is calculated and used
to derive a closed-form upper bound on the \ac{AWEP} of the SL-MIMO
system, and this derived expression is then used to analyze the diversity
gain of the system. We use our analytical results to design sub-optimal
codebooks to minimize the error rate of the SL-MIMO system. Simulation
results in $4\times4$ and $6\times6$ \ac{MIMO} systems with 4-ary,
16-ary, and 64-ary constellations show that our proposed SL-MIMO
scheme outperforms competing approaches such as X- and Y-codes in
terms of system error rate performance. SL-MIMO has 5.6~dB advantage
compared to X-codes and 4.7~dB advantage compared to Y-codes in $6\times6$
\ac{MIMO} system with a 64-ary constellation.\acresetall{}
\end{abstract}

\begin{IEEEkeywords}
\Ac{MIMO}, precoding, code-domain non-orthogonal transmission, spatial
coding, joint \ac{MGF} of ordered eigenvalues.\acresetall{}
\end{IEEEkeywords}

\section{Introduction\label{sec:Introduction}}

\Ac{MIMO} systems represent a key technology in wireless communications,
as they can substantially improve both the reliability and the data
rate \cite{goldsmith2005wireless}. For a \ac{MIMO} system with channel
state information available at the transmitter, the precoding technique
based on \ac{SVD} of the channel matrix is optimal with respect to
the capacity and error rate performance measures \cite{vu2007mimo}.
The reason for this is that standard \ac{SVD} precoding decomposes
the \ac{MIMO} channel into multiple parallel independent eigen-channels
\cite{goldsmith2005wireless,telatar1999capacity}, and an appropriate
power allocation across the eigen-channels can optimize different
performance criteria \cite{stoica2002maximum,sampath2001generalized,scaglione2002optimal,rostaing2002minimum}.
Hence, \ac{ML} detection at the receiver can be performed independently
for each eigen-channel. The downside of the standard \ac{SVD} precoding,
where the power allocation is realized by a diagonal matrix, is that
the diversity gain is determined by the smallest eigenvalue; therefore,
there is a trade-off between the diversity gain and data rate. However,
it was shown in \cite{palomar2003joint} that a non-diagonal precoding
matrix can improve the performance of the \ac{MIMO} system in terms
of the average \ac{BER}. Another non-diagonal precoding approach
was proposed in \cite{collin2004optimal} for a system with only two
eigen-channels, based on maximizing the minimum distance of the received
symbols. Motivated by \cite{collin2004optimal}, non-diagonal \ac{MIMO}
precoding schemes based on pairing good and bad eigen-channels have
been proposed in \cite{vrigneau2008extension,mohammed2011mimo,srinath2011low}
for larger \ac{MIMO} systems to improve the \ac{BER} performance
through increasing the diversity gain of the system. In \cite{vrigneau2008extension},
the authors proposed a non-diagonal cross-form matrix as an online
precoder\footnote{The terms ``precoder'' and ``code'' are defined in \cite{mohammed2011mimo}
to distinguish between ``online'' and ``offline'' design of the
precoding matrices. A \emph{precoder} is dedicated to each channel
realization and needs to be adaptively changed, while a \emph{code}
can be designed offline \emph{a~priori}.} with the aim of maximizing the minimum Euclidean distance within
pairs of eigen-channels. The so-called X- and Y-codes/precoders were
proposed in \cite{mohammed2011mimo}. X-codes/precoders use rotated
constellations within a pair of eigen-channels in order to minimize
the \ac{AWEP}, while Y-codes/precoders were designed to maximize
the minimum Euclidean distance by performing rate and power allocation
within a pair of eigen-channels. Another cross-form precoder design
was proposed in \cite{srinath2011low} to improve the error rate performance
while allowing low-complexity decoding. Despite the efficient design
of these precoders, all of the aforementioned \emph{pairing} approaches
suffer from interference within each pair of eigen-channels, specifically
at higher data rates. Besides, they need to be adapted to each channel
realization to perform more efficiently, since all of these precoders,
other than Y-codes, do not consider the stochastic properties of the
eigenvalues of the channel matrix.

On the other hand, code-domain non-orthogonal transmission has been
considered as a promising technology within modern wireless communications
to improve the spectral efficiency by sharing (re-using) the same
\acp{RE} \cite{dai2018survey,wang2018non,vaezi2019non,yuan20205g}.
\Ac{LDS} \cite{hoshyar2008novel,hoshyar2010lds} and \ac{SCMA} \cite{nikopour2013sparse}
are two prominent examples of this approach. In \ac{LDS}, each user
sparsely spreads its data over the available \acp{RE}. Therefore,
the interference pattern can be represented by a low-density factor
graph, and a corresponding low-complexity \ac{MP} detector can be
implemented at the receiver. The basic structure of \ac{SCMA} is
similar to that of \ac{LDS}, however, in \ac{SCMA}, instead of spreading
a symbol, each user transmits different symbols over its available
\acp{RE}. The potential of \ac{SCMA} has motivated researchers to
investigate the application of non-orthogonal transmission to \ac{MIMO}
systems. In this regard, various \emph{\ac{MIMO}-\ac{SCMA}} system
designs have been developed \cite{abdessamad2016performance,han2016uplink,liu2018spatial,pan2018uplink,elkawafi2019performance,liu2020spatial,pan2019multi,liu2021space}.
In \cite{abdessamad2016performance}, the authors analyzed the performance
of a multi-layer \ac{SMX} \ac{MIMO} system combined with \ac{SCMA}.
The performance of \ac{MIMO}-\ac{SCMA} systems designed based on
\ac{VBLAST} and \ac{STBC} was investigated in \cite{han2016uplink}.
The concept of \ac{SM} has also been considered in \ac{MIMO}-\ac{SCMA}
systems in \cite{liu2018spatial,pan2018uplink,elkawafi2019performance,liu2020spatial}
and the corresponding error rate performance of these systems has
been analyzed. The authors in \cite{pan2019multi} proposed a \ac{MSTBC}
\ac{MIMO}-\ac{SCMA} system to exploit the transmit diversity. In
\cite{liu2021space}, the concept of \ac{GSM} was used in a \ac{MIMO}-\ac{SCMA}
system utilizing \ac{STBC} in order to provide a framework for uplink
transmission to increase the transmit diversity and reduce the hardware
complexity of mobile users. In addition, various detection techniques
have been proposed to reduce the complexity and improve the performance
of the \ac{MIMO}-\ac{SCMA} system \cite{du2016joint,tang2016low,sharma2020low,pan2021low}.
However, in all of the aforementioned studies, code-domain non-orthogonal
transmission have been applied to either the \emph{time} or \emph{frequency}
domain, similar to the case of \ac{SISO}-\ac{SCMA}, and the concept
of non-orthogonal transmission in the \emph{spatial domain} has not
yet been applied to design \ac{MIMO} precoding. In addition to the
above-mentioned advantages of precoding, sharing and re-using the
eigen-channels can further enhance the spectral efficiency and error
rate performance of a \ac{MIMO} system.

Against this background, in this paper we introduce a new paradigm
for \ac{MIMO} precoding based on applying code-domain non-orthogonal
transmission across eigen-channels. In particular, the structure of
the code is designed to carefully control the interference among eigen-channels.
The contributions of this paper are as follows:
\begin{itemize}
\item Inspired by code-domain non-orthogonal transmission, we propose a
novel \ac{MIMO} scheme which uses \ac{SVD} precoding along with
a sparse non-diagonal matrix to allow re-use of the eigen-channels,
which we call \emph{sparse layered MIMO} (SL-MIMO). We exploit \ac{SVD}
precoding to create a set of eigen-channels which we use as orthogonal
\acp{RE}. Then, we split the input data bits into layers and spread
the corresponding symbols over the eigen-channels in a sparse manner.
\item We propose a sub-optimal \ac{MP} detection algorithm for SL-MIMO,
which has a significantly lower decoding complexity than \ac{ML}
detection. We show that the \ac{MP} detector performance exhibits
fast convergence and has a performance very close to that of \ac{ML}.
\item We analyze the \ac{AWEP} of the SL-MIMO system, and derive a closed-form
upper bound on the \ac{AWEP} which is tight at high \ac{SNR}. This
requires evaluation of the joint \ac{MGF} of the \emph{ordered}
eigenvalues, which to the best of the authors' knowledge has not been
previously provided in the literature. Moreover, we perform an asymptotic
analysis of the \ac{AWEP} of the proposed system and determine its
diversity gain. 
\item A sub-optimal approach is proposed for designing the sparse codebooks
for the layers, in order to minimize the \ac{AWEP} of the system.
We formulate a joint optimization problem to determine the multidimensional
constellations for all layers. We then define a multi-step sub-optimal
approach to simplify the procedure. In each step, we aim to minimize
the asymptotic \ac{AWEP} for a set of layers. The multi-step design
procedure has a low complexity and only needs to be performed once
offline. In this way, the SL-MIMO approach is categorized as a ``code''
and not as a ``precoder'' \cite{mohammed2011mimo}, i.e., the constellations
are designed \emph{a~priori} and do not change with each channel
realization.
\item Finally, we investigate the performance of the system through numerical
simulations. The results show substantial improvement over competing
approaches such as X- and Y-codes, especially at higher data rates.
The superiority increases at higher data rates and in larger \ac{MIMO}
systems.
\end{itemize}
The rest of this paper is organized as follows. We describe the SL-MIMO
system model in Section~\ref{sec:System-Model}. In Section~\ref{sec:Performance-Analysis},
the analysis of the \ac{AWEP}, including its asymptotic analysis,
is presented. Section~\ref{sec:Constellation-Design} describes the
constellation design procedure for the SL-MIMO system. Numerical
results and comparisons with other relevant schemes are provided in
Section~\ref{sec:Numerical-Results}. Finally, Section~\ref{sec:Conclusion}
concludes this paper.

\emph{Notation:} Boldface lower-case letters indicate column vectors,
and boldface upper-case letters indicate matrices. $\mathbf{I}_{n}$
represents an $n\times n$ identity matrix, and $\mathbf{0}_{n\times m}$
indicates an $n\times m$ zero matrix. The superscripts $\left(\cdot\right)^{T}$
and $\left(\cdot\right)^{H}$ denote transpose and Hermitian transpose,
respectively. $\mbox{diag}\left(\cdot\right)$ converts a vector into
a corresponding diagonal matrix. $\left\Vert \cdot\right\Vert $ denotes
the Euclidean norm of a vector, and $\left|\cdot\right|$ denotes
the cardinality of a set. The set $\left\{ 1,2,\dots,K\right\} $
is represented by $[K]$. $\mathsf{\mathbb{E}}\left\{ \cdot\right\} $
denotes the expectation operator. Finally, the set of complex numbers
is denoted by $\mathbb{C}$.

\section{System Model\label{sec:System-Model}}

\begin{figure*}[t]
\begin{centering}
\includegraphics[scale=0.4]{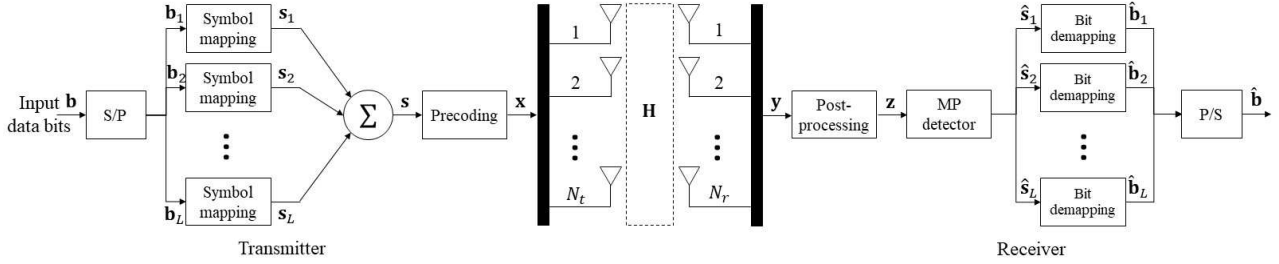}
\par\end{centering}
\caption{Transmitter and receiver structure for the proposed \acf{SL-MIMO}
system.\label{fig:A-schematic-of SL-MIMO}}
\end{figure*}
Fig.~\ref{fig:A-schematic-of SL-MIMO} shows a block diagram of the
proposed SL-MIMO system. We consider a point to point communication
system, whose transmitter and receiver employ $N_{t}$ and $N_{r}$
antennas, respectively. The elements of the channel matrix $\mathbf{H}\in\mathbb{C}^{N_{r}\times N_{t}}$
are i.i.d complex Gaussian random variables with zero mean and unit
variance. The received signal is given by 
\begin{equation}
\mathbf{\mathbf{y}}=\mathbf{H}\mathbf{x}+\mathbf{n},\label{eq:y=00003DHx+n}
\end{equation}
where $\mathbf{x}\in\mathbb{C}^{N_{t}\times1}$ is the \emph{transmit vector},
and $\mathbf{n}\in\mathbb{C}^{N_{r}\times1}$ is the noise vector
which is distributed according to $\mathcal{CN}\left(0,N_{0}\mathbf{I}_{N_{r}}\right)$.
We use the \ac{SVD} precoding technique to decompose the channel
into a set of parallel virtual channels (or eigen-channels). The \ac{SVD}
of the channel matrix $\mathbf{H}$ can be expressed as $\mathbf{H}=\mathbf{U}\boldsymbol{\Sigma}\mathbf{V}^{H}$,
where $\mathbf{U}\in\mathbb{C}^{N_{r}\times N_{r}}$ and $\mathbf{V}\in\mathbb{C}^{N_{t}\times N_{t}}$
are unitary matrices, and $\boldsymbol{\Sigma}=\left[\boldsymbol{\Lambda}^{\frac{1}{2}},\mathbf{0}_{N_{r}\times(N_{t}-N_{r})}\right]$
for $N_{r}<N_{t}$, or $\boldsymbol{\Sigma}=\left[\boldsymbol{\Lambda}^{\frac{1}{2}},\mathbf{0}_{N_{t}\times(N_{r}-N_{t})}\right]^{T}$
for $N_{r}\geq N_{t}$, where $\boldsymbol{\Lambda}=\mbox{diag}\left(\boldsymbol{\lambda}=\left[\lambda_{1},\dots,\lambda_{n}\right]^{T}\right)$
and $\left\{ \lambda_{m}\right\} $ are the eigenvalues of the matrix
$\mathbf{H}\mathbf{H}^{H}$ ($N_{r}<N_{t}$) or $\mathbf{H}^{H}\mathbf{H}$
($N_{r}\geq N_{t}$), such that $n=\min\left(N_{t},N_{r}\right)$
and $\lambda_{1}\geq\lambda_{2}\geq\dots\geq\lambda_{n}\geq0$. \Ac{SVD}
transmission implies that at the transmitter, the \emph{data symbol vector}
$\mathbf{s}=\left[s_{1},\dots,s_{n}\right]^{T}$ is mapped to the
transmit vector $\mathbf{x}$ via $\mathbf{x}=\bar{\mathbf{V}}\mathbf{s}$,
while at the receiver, the received vector $\mathbf{y}$ is pre-multiplied
by the matrix $\bar{\mathbf{U}}^{H}$, where $\bar{\mathbf{V}}$ and
$\bar{\mathbf{U}}$ are sub-matrices consisting of the first $n$
columns of $\mathbf{V}$ and $\mathbf{U}$, respectively. Then, the
post-processed received signal vector is given as 
\begin{align}
\mathbf{z} & \triangleq\bar{\mathbf{U}}^{H}\mathbf{y}=\boldsymbol{\Lambda}^{\frac{1}{2}}\mathbf{s}+\mathbf{n}',\label{eq:z=00003DLambda s + n}
\end{align}
where $\mathbf{n}'\in\mathbb{C}^{n\times1}$ is the post-processed
noise vector which has the same distribution as $\mathbf{n}$. The
system in (\ref{eq:z=00003DLambda s + n}) consists of $n$ parallel
eigen-channels with channel gains $\left\{ \sqrt{\lambda_{m}}\right\} $,
which are used as orthogonal resources to transmit sparse coded symbols.

In the proposed SL-MIMO system, each block of $\eta=L\times\log_{2}M$
bits in the data bit stream $\mathbf{b}$ is split into $L$ equal-sized
packets $\mathbf{b}_{l}$; each $\log_{2}M$-bit packet $\mathbf{b}_{l}$
represents the data bits for layer $l$, and is encoded into a sparse
codeword $\mathbf{s_{\mathit{l}}}=\left[s_{l,1},\dots,s_{l,n}\right]^{T}$
of length $n$. The codeword $\mathbf{s}_{l}$ is selected from a
pre-designed codebook $\mathcal{S}_{l}$ for layer $l$, having size
$M$ (for the sake of simplicity, we assume that the codebooks of
all layers are equal in size). Thus, the system achieves a spectral
efficiency of $\eta$ \ac{bpcu}.

Each codebook $\mathcal{S}_{l}$ is \emph{sparse}, in the sense that
only $n_{l}$ of the $n$ entries of any codeword consist of nonzero
\ac{IQ} symbols; the remainder of the entries are equal to zero.
The set of nonzero positions for codebook $\mathcal{S}_{l}$ is denoted
by $\mathcal{N}_{l}\subset\{1,2,\ldots,n\}$, and so we have $|\mathcal{N}_{l}|=n_{l}$.
Thus, for $m\in\mathcal{N}_{l}$, entry $s_{l,m}$ in $\mathbf{s}_{l}$
is then the \ac{IQ} symbol for layer $l$ which is transmitted over
eigen-channel $m$. The number of layers that share the $m$-th eigen-channel
is denoted by $d_{m}$, and $\mathcal{D}_{m}\subset\{1,2,\ldots,L\}$
is the set of the interfering layers at eigen-channel $m$. The connectivity
between the eigen-channels and the layers can be represented by a
binary \ac{SL} matrix (similar to the parity-check matrix of a low-density
parity-check (LDPC) code) whose rows corresponds to eigen-channels
and columns corresponds to layers; its $(m,l)$-entry $a_{m,l}$ is
equal to 1 if and only if layer $l$ uses eigen-channel $m$.

As an example, we consider the \ac{SL} matrix 
\begin{equation}
\mathbf{A}=\left[\begin{array}{cccccc}
1 & 1 & 1 & 0 & 0 & 0\\
1 & 0 & 0 & 1 & 1 & 0\\
0 & 1 & 0 & 1 & 0 & 1\\
0 & 0 & 1 & 0 & 1 & 1
\end{array}\right]\label{eq:SL matrix A}
\end{equation}
which corresponds to the case where $n=4$ and $L=6$. It can be deduced
from the first column of this matrix that layer 1 uses eigen-channels
1 and 2, i.e., $\mathcal{N}_{1}=\left\{ 1,2\right\} $, and that eigen-channel
1 is used by layers 1, 2 and 3, i.e., $\mathcal{D}_{1}=\left\{ 1,2,3\right\} $.
We also note that the matrix $\mathbf{A}$ in (\ref{eq:SL matrix A})
is a \emph{regular} matrix, which means that the number of ones in
rows and in columns are fixed for all eigen-channels and layers, respectively.
In other words $n_{l}=2$ for all $l$, and $d_{m}=3$ for all $m$.

The length-$n$ data symbol vector is then formed by adding the codewords
for each layer, i.e., 
\begin{equation}
\mathbf{s}=\sum_{l=1}^{L}\mathbf{s}_{l},\label{eq:transmit symbol}
\end{equation}
so that we have 
\begin{equation}
s_{m}=\sum_{l\in\mathcal{D}_{m}}s_{l,m},\;\forall m,\label{eq:}
\end{equation}
Then, we define $\mathcal{S}$ to be set of all possible data symbol
vectors $\mathbf{s}$. In addition, we impose the following design
condition on the codebooks $\left\{ \mathcal{S}_{l}\right\} $: we
assume that the codebooks satisfy the property that any two vectors
in $\mathcal{S}$ differ in some position $m$ only if there exists
at least one layer sharing eigen-channel $m$ in which the chosen
codewords also differ, i.e., for any vectors $\mathbf{s}^{(1)}=\sum_{l=1}^{L}\mathbf{s}_{l}^{(1)}$
and $\mathbf{s}^{(2)}=\sum_{l=1}^{L}\mathbf{s}_{l}^{(2)}$, it holds
that for every $m\in\{1,2,\ldots,n\}$

\begin{equation}
s_{m}^{(1)}\ne s_{m}^{(2)}\quad\mbox{only if}\quad\mathbf{s}_{l}^{(1)}\ne\mathbf{s}_{l}^{(2)}\quad\mbox{for at least one}\;l\in\mathcal{D}_{m}.\label{eq:new condition}
\end{equation}

Also, for ease of exposition, we assume that all codebooks have the
same average energy $E_{s}$, i.e.,
\begin{equation}
\mathsf{\mathbb{E}}\left\{ \left\Vert \mathbf{s}_{l}\right\Vert ^{2}\right\} =E_{s},\ l=1,\dots,L.\label{eq:Energy of layers}
\end{equation}
Finally, we assume that input data bits of different layers are independent,
so that the total transmit power of the system is
\[
P_{t}=\mathsf{\mathbb{E}}\left\{ \left\Vert \mathbf{x}\right\Vert ^{2}\right\} =\mathsf{\mathbb{E}}\left\{ \left\Vert \mathbf{s}\right\Vert ^{2}\right\} =\mathsf{\mathbb{E}}\left\{ \left\Vert \sum_{l=1}^{L}\mathbf{s}_{l}\right\Vert ^{2}\right\} =LE_{s},
\]
where the second equality holds since $\bar{\mathbf{V}}^{H}\bar{\mathbf{V}}=\mathbf{I}_{n}$,
and the final equality holds due to the aforementioned independence
assumption. Hence, the SNR is equal to $P_{t}/N_{0}$.

With these properties, the SL-MIMO system can be considered as a
generalized \ac{SVD} precoding system, with additional degrees of
freedom. Also, it is worthwhile to note that the X- and Y-codes of
\cite{mohammed2011mimo} may be viewed as special cases of SL-MIMO.
The corresponding \ac{SL} matrices of X- and Y-codes have the shape
of the letters X and Y, respectively. For example, the \ac{SL} matrices
of X- and Y-codes for a system with 4 eigen-channels are respectively
given by
\[
\mathbf{A}_{\mathrm{X}}=\left[\begin{array}{cccc}
1 & 0 & 0 & 1\\
0 & 1 & 1 & 0\\
0 & 1 & 1 & 0\\
1 & 0 & 0 & 1
\end{array}\right],\mathbf{A}_{\mathrm{Y}}=\left[\begin{array}{cccc}
1 & 0 & 0 & 1\\
0 & 1 & 1 & 0\\
0 & 1 & 0 & 0\\
1 & 0 & 0 & 0
\end{array}\right].
\]
The structure of X- and Y-codes essentially minimizes overlap between
eigen-channels, and allows for separate (independent) detection on
each pair of eigen-channels. On the other hand, we allow for a controlled
(and carefully designed) overlap between eigen-channels, which results
in an \emph{interference diversity} that can improve the performance,
while we use \ac{MP} detection in order to keep the receiver complexity
low.

\subsection{Optimal (Maximum Likelihood) Detection\label{subsec:Optimal-(Maximum-Likelihood)-Det}}

At the receiver, the optimum \acf{ML} detector is given by
\begin{equation}
\hat{\mathbf{s}}=\arg\underset{\mathbf{s}\in\mathcal{S}}{\min}\left\{ \left\Vert \mathbf{z}-\boldsymbol{\Lambda}^{\frac{1}{2}}\mathbf{s}\right\Vert ^{2}\right\} .\label{eq:ML}
\end{equation}
Recalling (\ref{eq:transmit symbol}), we have $\left|\mathcal{S}\right|=\prod_{l=1}^{L}\left|\mathcal{S}_{l}\right|=M^{L}$;
thus the complexity of \ac{ML} detection is $\mathcal{O}\left(M^{L}\right)$.

\subsection{Message-Passing Algorithm\label{subsec:Message-Passing-Algorithm}}

\Ac{ML} detection of SL-MIMO is too complex for practical implementation,
especially for large values of $M$ and $L$. Hence, in the following
we propose the use of a \acf{MP} algorithm for detection in the SL-MIMO
system, in which the receiver exploits the sparsity of the underlying
factor graph to detect the received symbols with a lower complexity
(the \ac{ML} detector will later be used as a benchmark to validate
the \ac{MP} detection results).

The decisions $\{\hat{\mathbf{s}}_{1},\hat{\mathbf{s}}_{2},\ldots,\hat{\mathbf{s}}_{L}\}$
are obtained by marginalizing the function $P(\mathbf{s}_{1},\mathbf{s}_{2},\ldots,\mathbf{s}_{L}|\mathbf{z})$,
which is proportional to 
\begin{equation}
p(\mathbf{z}|\mathbf{s}_{1},\mathbf{s}_{2},\ldots,\mathbf{s}_{L})P(S)=\prod_{m=1}^{n}P(z_{m}|S_{m})\prod_{l=1}^{L}P(\mathbf{s}_{l}),\label{eq:p(z|=00007Bs_l=00007D)}
\end{equation}
where $S$ denotes the collection of all variables $\mathbf{s}_{l}$,
and $S_{m}$ denotes the collection of all variables $\mathbf{s}_{l}$
for $l\in\mathcal{D}_{m}$. The factor graph then has a variable node
for each variable $\mathbf{s}_{l}$, and a factor node $z_{m}$ corresponding
to each factor 
\begin{equation}
P(z_{m}|S_{m})=\frac{1}{\pi N_{0}}\exp\left(-\frac{1}{N_{0}}\left|z_{m}-\sqrt{\lambda_{m}}\sum_{l\in\mathcal{D}_{m}}s_{l,m}\right|^{2}\right).\label{eq:p(z_m|s_m)}
\end{equation}
The \ac{MP} algorithm can then be described as follows. The algorithm
is initialized via 
\[
\mu_{s_{l}\rightarrow z_{m}}^{0}(\mathbf{s}_{l})=1\mbox{ for all }m,l\in\mathcal{D}_{m},\mbox{and }\mathbf{s}_{l}\in\mathcal{S}_{l}.
\]
Then, \ac{MP} detection proceeds as follows, where $N$ denotes the
total number of iterations. For each $t=1$ to $N$, compute 
\begin{equation}
\mu_{z_{m}\rightarrow\mathbf{s}_{l}}^{t}(\mathbf{s}_{l})=\sum_{\{\mathbf{s}_{l'}\},l'\in\mathcal{D}_{m}\backslash\{l\}}P(z_{m}|S_{m})\prod_{l'\in\mathcal{D}_{m}\backslash\{l\}}\mu_{\mathbf{s}_{l'}\rightarrow z_{m}}^{t-1}(\mathbf{s}_{l'})\label{eq:Mu(z_m to s_l)}
\end{equation}
 and 
\begin{equation}
\mu_{\mathbf{s}_{l}\rightarrow z_{m}}^{t}(\mathbf{s}_{l})=\prod_{m'\in\mathcal{N}_{l}\backslash\{m\}}\mu_{z_{m'}\rightarrow\mathbf{s}_{l}}^{t}(\mathbf{s}_{l})P(\mathbf{s}_{l})\label{eq:Mu(s_l to z_m)}
\end{equation}
which, introducing a scaling factor, can be rewritten as 
\begin{equation}
\mu_{\mathbf{s}_{l}\rightarrow z_{m}}^{t}(\mathbf{s}_{l})=\prod_{m'\in\mathcal{N}_{l}\backslash\{m\}}\mu_{z_{m'}\rightarrow\mathbf{s}_{l}}^{t}(\mathbf{s}_{l})\label{eq:Mu(s_l to z_m)-2}
\end{equation}
 for $\mathbf{s}_{l}\in\mathcal{S}_{l}$, and $\mu_{z_{m}\rightarrow\mathbf{s}_{l}}^{t}(\mathbf{s}_{l})=0$
otherwise. Finally, decisions are made via 
\begin{equation}
\hat{\mathbf{s}}_{l}=\arg\max_{\mathbf{s}_{l}\in\mathcal{S}_{l}}P(\mathbf{s}_{l})\prod_{m\in\mathcal{N}_{l}}\mu_{z_{m}\rightarrow\mathbf{s}_{l}}^{N}(\mathbf{s}_{l}).\label{eq:MP decision}
\end{equation}

The \ac{MP} detector exploits the \emph{interference diversity}
among the layers and eigen-channels. This diversity helps the \ac{MP}
detector to efficiently update the messages in each iteration. The
complexity of the \ac{MP} detector is dominated by (\ref{eq:Mu(z_m to s_l)}),
where the \emph{a~posteriori} probabilities of $\left\{ z_{m}\right\} $
are computed, and its complexity depends on the number of interfering
layers at the eigen-channels. Therefore, the overall complexity of
the \ac{MP} detector is $\mathcal{O}\left(M^{d_{\max}}\right)$,
where $d_{\max}=\underset{m}{\max}\left\{ d_{m}\right\} $.

\section{Performance Analysis\label{sec:Performance-Analysis}}

In this section, we analyze the \acf{AWEP} of the proposed SL-MIMO
system. This \ac{AWEP} analysis will later be used for codebook design
in Section~\ref{sec:Constellation-Design}. Our analysis will focus
on \ac{ML} detection, as this is more tractable than that of the
\ac{MP} detector and also provides an upper bound on the achievable
performance of the \ac{MP} detector; moreover, as we will show later
in Section~\ref{sec:Numerical-Results}, this upper bound becomes
very tight at high SNR. The \ac{AWEP} can be calculated by finding
the \acf{PEP} associated with data symbol vectors $\mathbf{s}$ and
$\hat{\mathbf{s}}$:
\begin{align}
\mathrm{PEP}\left(\mathbf{s},\hat{\mathbf{s}}\right) & =\Pr\left(\mathbf{s}\rightarrow\hat{\mathbf{s}}|\boldsymbol{\Lambda}\right)\nonumber \\
 & =\Pr\left(\left\Vert \mathbf{z}-\boldsymbol{\Lambda}^{\frac{1}{2}}\mathbf{s}\right\Vert ^{2}>\left\Vert \mathbf{z}-\boldsymbol{\Lambda}^{\frac{1}{2}}\mathbf{\hat{s}}\right\Vert ^{2}\right)\nonumber \\
 & =\Pr\left(\left\Vert \boldsymbol{\Lambda}^{\frac{1}{2}}\left(\mathbf{s}-\mathbf{\hat{s}}\right)\right\Vert ^{2}<2\Re\left\{ \left(\mathbf{s}-\mathbf{\hat{s}}\right)^{H}\boldsymbol{\Lambda}^{\frac{1}{2}}\mathbf{n}^{\prime}\right\} \right)\nonumber \\
 & =\mathrm{Q}\left(\sqrt{\frac{\left\Vert \boldsymbol{\Lambda}^{\frac{1}{2}}\left(\mathbf{s}-\mathbf{\hat{s}}\right)\right\Vert ^{2}}{2N_{0}}}\right)\nonumber \\
 & =\mathrm{Q}\left(\sqrt{\frac{\sum_{m=1}^{n}\lambda_{m}\left|\psi_{m}\right|^{2}}{2N_{0}}}\right),\label{eq:Q function}
\end{align}
where $\boldsymbol{\psi}=\left[\psi_{1},\psi_{2},\dots,\psi_{n}\right]^{T}=\mathbf{s}-\hat{\mathbf{s}}$
is the \emph{difference vector} for this symbol vector pair. The
evaluation of the average \ac{PEP} requires use of the joint \ac{PDF}
of the eigenvalues. The joint \ac{PDF} of the \emph{ordered} eigenvalues
$\lambda_{1}\geq\lambda_{2}\dots\geq\lambda_{n-1}\geq\lambda_{n}$
is given by \cite{chiani2003capacity}
\begin{equation}
p\left(\boldsymbol{\lambda}\right)=\frac{1}{C_{N_{t},N_{r}}}\prod_{m}e^{-\lambda_{m}}\lambda_{m}^{\left|N_{t}-N_{r}\right|}\prod_{m>m'}\left(\lambda_{m}-\lambda_{m'}\right)^{2},\label{eq:PDF}
\end{equation}
 where $C_{N_{t},N_{r}}$ is a normalizing factor. After expanding
the polynomial, it can be written as 
\begin{equation}
p\left(\boldsymbol{\lambda}\right)=\frac{1}{C_{N_{t},N_{r}}}\sum_{p=1}^{P}\alpha_{p}\prod_{m}e^{-\lambda_{m}}\lambda_{m}^{\beta_{p,m}},\label{eq:pdf}
\end{equation}
 where $P$ indicates the number of the terms in $p\left(\boldsymbol{\lambda}\right)$,
and $\alpha_{p}$ and $\beta_{p,m}$ are the coefficients of the $p$-th
monomial and the exponent of $\lambda_{m}$ in that monomial, respectively.

Using the upper bound of $Q$-function \cite{chiani2003new}
\begin{equation}
Q\left(x\right)\leq\frac{1}{12}e^{-\frac{x^{2}}{2}}+\frac{1}{4}e^{-\frac{2x^{2}}{3}},\label{eq:Approximate Q function}
\end{equation}
the average \ac{PEP} can be expressed as
\begin{align}
\overline{\mathrm{PEP}}\mathrm{\left(\mathbf{s},\hat{\mathbf{s}}\right)}= & \mathsf{\mathbb{E}}_{\boldsymbol{\lambda}}\left\{ \mathrm{Q}\left(\sqrt{\frac{\sum_{m}\lambda_{m}\left|\psi_{m}\right|^{2}}{2N_{0}}}\right)\right\} \nonumber \\
\leq & \mathsf{\mathbb{E}}_{\boldsymbol{\lambda}}\left\{ \frac{1}{12}\prod_{m}e^{-\frac{\lambda_{m}\left|\psi_{m}\right|^{2}}{4N_{0}}}+\frac{1}{4}\prod_{m}e^{-\frac{\lambda_{m}\left|\psi_{m}\right|^{2}}{3N_{0}}}\right\} \nonumber \\
= & \frac{1}{12}M_{\boldsymbol{\lambda}}\left(-\mathbf{a}\right)+\frac{1}{4}M_{\boldsymbol{\lambda}}\left(-\mathbf{b}\right)\nonumber \\
= & \underset{I_{1}\left(\boldsymbol{\psi}\right)}{\underbrace{\frac{1}{12}\dotsint_{\mathcal{D}_{ord}}\prod_{m}e^{-\frac{\lambda_{m}\left|\psi_{m}\right|^{2}}{4N_{0}}}p\left(\boldsymbol{\lambda}\right)d\boldsymbol{\lambda}}}\nonumber \\
 & +\underset{I_{2}\left(\boldsymbol{\psi}\right)}{\underbrace{\frac{1}{4}\dotsint_{\mathcal{D}_{ord}}\prod_{m}e^{-\frac{\lambda_{m}\left|\psi_{m}\right|^{2}}{3N_{0}}}p\left(\boldsymbol{\lambda}\right)d\boldsymbol{\lambda}}},\label{eq:average PEP}
\end{align}
where $\mathbf{a}=\left[a_{1},a_{2},\dots,a_{n}\right]^{T}$ and $a_{m}=\frac{\left|\psi_{m}\right|^{2}}{4N_{0}}$
for each $m$, $\mathbf{b}=\left[b_{1},b_{2},\dots,b_{n}\right]^{T}$
and $b_{m}=\frac{\left|\psi_{m}\right|^{2}}{3N_{0}}$ for each $m$
(note that each coefficient $a_{m}$ and $b_{m}$ is proportional
to the \ac{SNR}), $M_{\boldsymbol{\lambda}}\left(\boldsymbol{\chi}\right)=\mathbb{E}_{\boldsymbol{\lambda}}\left\{ \prod_{m=1}^{n}e^{\chi_{m}\lambda_{m}}\right\} $
is the joint \ac{MGF} of the eigenvalues, and the multiple integral
is over the domain $\mathcal{D}_{ord}=\left\{ \infty>\lambda_{1}\geq\lambda_{2}\dots\geq\lambda_{n-1}\geq\lambda_{n}\geq0\right\} $.
Therefore, to compute the average \ac{PEP}, calculation of the joint
moments of the \emph{ordered} eigenvalues is required.

Expanding the first term in (\ref{eq:average PEP}), which we denote
by $I_{1}\left(\boldsymbol{\psi}\right)$, we have
\begin{align}
I_{1}\left(\boldsymbol{\psi}\right)\triangleq & \frac{1}{12}\dotsint_{\mathcal{D}_{ord}}\prod_{m}e^{-a_{m}\lambda_{m}}p\left(\boldsymbol{\lambda}\right)d\boldsymbol{\lambda}\nonumber \\
= & \frac{1}{12C_{N_{t},N_{r}}}\sum_{p=1}^{P}\alpha_{p}\dotsint_{\mathcal{D}_{ord}}\prod_{m}e^{-\left(1+a_{m}\right)\lambda_{m}}\lambda_{m}^{\beta_{p,m}}d\boldsymbol{\lambda}\nonumber \\
= & \frac{1}{12C_{N_{t},N_{r}}}\sum_{p=1}^{P}\alpha_{p}\int_{0}^{\infty}e^{-\left(1+a_{n}\right)\lambda_{n}}\lambda_{n}^{\beta_{p,n}}\nonumber \\
 & \times\int_{\lambda_{n}}^{\infty}e^{-\left(1+a_{n-1}\right)\lambda_{n-1}}\lambda_{n-1}^{\beta_{p,n-1}}\times\dots\nonumber \\
 & \times\int_{\lambda_{2}}^{\infty}e^{-\left(1+a_{1}\right)\lambda_{1}}\lambda_{1}^{\beta_{p,1}}d\lambda_{1}\dots d\lambda_{n-1}d\lambda_{n},\label{eq:int seq}
\end{align}
this multiple integral will be solved sequentially. The inner integral
in (\ref{eq:int seq}) is in the form of the upper incomplete Gamma
function defined as \cite[eq. 6.5.3]{abramowitz1972handbook}
\begin{equation}
\Gamma\left(a,x\right)=\int_{x}^{\infty}e^{-t}t^{a-1}dt.\label{eq:Gamma def}
\end{equation}
Note that for an integer $a\geq1$, we have
\begin{equation}
\Gamma\left(a,x\right)=\left(a-1\right)!e^{-x}\sum_{i=0}^{a-1}\frac{x^{i}}{i!}.\label{eq:Gamma Solution}
\end{equation}
Hence the inner integral can be evaluated as
\begin{align}
 & \int_{\lambda_{2}}^{\infty}e^{-\left(1+a_{1}\right)\lambda_{1}}\lambda_{1}^{\beta_{p,1}}d\lambda_{1}\nonumber \\
 & =\frac{\beta_{p,1}!}{\left(1+a_{1}\right)^{\beta_{p,1}+1}}e^{-\left(1+a_{1}\right)\lambda_{2}}\sum_{i_{1}=0}^{\beta_{p,1}}\frac{\left(\left(1+a_{1}\right)\lambda_{2}\right)^{i_{1}}}{i_{1}!}.\label{eq:inner int}
\end{align}
After substituting (\ref{eq:inner int}) into (\ref{eq:int seq}),
it can be observed that the next integral has the structure of a series
each of whose terms corresponds to an upper incomplete Gamma function.
Similar forms emerge in the solution of the next integral over $\lambda_{2}$.
Continuing in this manner, after calculating the sequence of $\left(n-1\right)$
integrals in the form of (\ref{eq:Gamma def}), we finally have a
series of integrals over $\lambda_{n}$ each of which has the structure
of the Gamma function defined as \cite[eq. 6.1.1]{abramowitz1972handbook}
$\Gamma\left(a\right)=\int_{0}^{\infty}e^{-t}t^{a-1}dt$; also, it
is known that for an integer $a\geq1$, $\Gamma\left(a\right)=\left(a-1\right)!$.
Hence, $I_{1}\left(\boldsymbol{\psi}\right)$ can be solved as (\ref{eq:int I1}),{\footnotesize{}}
\begin{figure*}[tbh]
\begin{align}
I_{1}\left(\boldsymbol{\psi}\right)= & \frac{1}{12C_{N_{t},N_{r}}}\sum_{p=1}^{P}\biggl(\alpha_{p}\left(1\right)_{\beta_{p,1}}\frac{1}{\left(1+a_{1}\right)^{\beta_{p,1}+1}}\sum_{i_{1}=0}^{\beta_{p,1}}\biggl(\left(i_{1}+1\right)_{\beta_{p,2}}\frac{\left(1+a_{1}\right)^{i_{1}}}{\left(2+a_{1}+a_{2}\right)^{\beta_{p,2}+i_{1}+1}}\sum_{i_{2}=0}^{\beta_{p,2}+i_{1}}\biggl(\left(i_{2}+1\right)_{\beta_{p,3}}\nonumber \\
 & \times\frac{\left(2+a_{1}+a_{2}\right)^{i_{2}}}{\left(3+a_{1}+a_{2}+a_{3}\right)^{\beta_{p,3}+i_{2}+1}}\dots\sum_{i_{n-1}=0}^{\beta_{p,n-1}+i_{n-2}}\biggl(\left(i_{n-1}+1\right)_{\beta_{p,n}}\times\frac{\left(n-1+a_{1}+\dots+a_{n-1}\right)^{i_{n-1}}}{\left(n+a_{1}+\dots+a_{n}\right)^{\beta_{p,n}+i_{n-1}+1}}\biggr)\biggr)\biggr)\biggr).\label{eq:int I1}\\
\hline \nonumber 
\end{align}
\end{figure*}
 where $\left(x\right)_{k}=\frac{\left(x+k-1\right)!}{\left(x-1\right)!}$
denotes the rising factorial, and there are $(n-1)$ nested summations.
The second term in (\ref{eq:average PEP}), denoted by $I_{2}\left(\boldsymbol{\psi}\right)$,
can be evaluated in the same manner. Finally, the \ac{AWEP} can be
derived according to the union bound as \cite{proakis1995digital}
\begin{align}
\mathrm{AWEP} & \leq\frac{1}{2^{\eta}}\sum_{\mathbf{s}}\sum_{\hat{\mathbf{s}}}\overline{\mathrm{PEP}}\left(\mathbf{s},\hat{\mathbf{s}}\right)\nonumber \\
 & \leq\frac{1}{2^{\eta}}\sum_{\mathbf{s}}\sum_{\hat{\mathbf{s}}}\left(I_{1}\left(\boldsymbol{\psi}\right)+I_{2}\left(\boldsymbol{\psi}\right)\right).\label{eq:AWEP}
\end{align}
The complete expression is shown in (\ref{eq:complete AWEP}).
\begin{center}
{\footnotesize{}}
\begin{figure*}[tbh]
\centering{}
\begin{align}
\mathrm{AWEP}\leq & \frac{1}{2^{\eta}C_{N_{t},N_{r}}}\sum_{\mathbf{s}}\sum_{\hat{\mathbf{s}}}\sum_{p=1}^{P}\alpha_{p}\left[\left(\frac{1}{12}\frac{\left(1\right)_{\beta_{p,1}}}{\left(1+a_{1}\right)^{\beta_{p,1}+1}}\sum_{i_{1}=0}^{\beta_{p,1}}\frac{\left(i_{1}+1\right)_{\beta_{p,2}}\left(1+a_{1}\right)^{i_{1}}}{\left(2+a_{1}+a_{2}\right)^{\beta_{p,2}+i_{1}+1}}\right.\right.\nonumber \\
 & \left.\times\sum_{i_{2}=0}^{\beta_{p,2}+i_{1}}\frac{\left(i_{2}+1\right)_{\beta_{p,3}}\left(2+a_{1}+a_{2}\right)^{i_{2}}}{\left(3+a_{1}+a_{2}+a_{3}\right)^{\beta_{p,3}+i_{2}+1}}\dots\sum_{i_{n-1}=0}^{\beta_{p,n-1}+i_{n-2}}\frac{\left(i_{n-1}+1\right)_{\beta_{p,n}}\left(n-1+a_{1}+\dots+a_{n-1}\right)^{i_{n-1}}}{\left(n+a_{1}+\dots+a_{n}\right)^{\beta_{p,n}+i_{n-1}+1}}\right)\nonumber \\
 & +\left(\frac{1}{4}\frac{\left(1\right)_{\beta_{p,1}}}{\left(1+b_{1}\right)^{\beta_{p,1}+1}}\sum_{i_{1}=0}^{\beta_{p,1}}\frac{\left(i_{1}+1\right)_{\beta_{p,2}}\left(1+b_{1}\right)^{i_{1}}}{\left(2+b_{1}+b_{2}\right)^{\beta_{p,2}+i_{1}+1}}\sum_{i_{2}=0}^{\beta_{p,2}+i_{1}}\frac{\left(i_{2}+1\right)_{\beta_{p,3}}\left(2+b_{1}+b_{2}\right)^{i_{2}}}{\left(3+b_{1}+b_{2}+b_{3}\right)^{\beta_{p,3}+i_{2}+1}}\dots\right.\nonumber \\
 & \left.\left.\times\sum_{i_{n-1}=0}^{\beta_{p,n-1}+i_{n-2}}\frac{\left(i_{n-1}+1\right)_{\beta_{p,n}}\left(n-1+b_{1}+\dots+b_{n-1}\right)^{i_{n-1}}}{\left(n+b_{1}+\dots+b_{n}\right)^{\beta_{p,n}+i_{n-1}+1}}\right)\right].\label{eq:complete AWEP}\\
\hline \nonumber 
\end{align}
\end{figure*}
{\footnotesize\par}
\par\end{center}

\subsection*{Asymptotic Analysis\label{subsec:Asymptotic-Analysis}}

In this subsection, we will derive a high-\ac{SNR} approximate expression
for the \ac{AWEP}, and also analyze the diversity gain of the proposed
SL-MIMO system. To this end, first we define, for a vector $\boldsymbol{\psi}\neq\mathbf{0}$,
a function $\mathfrak{F}$ which counts the number of leading zeros
of $\boldsymbol{\psi}$, i.e.,
\begin{equation}
\mathfrak{F}\left(\boldsymbol{\psi}\right)=\begin{cases}
c-1 & \mathrm{if}\ \psi_{c}\ne0\ \mbox{and}\ \psi_{m}=0\ \forall m<c.\end{cases}\label{eq:Function F}
\end{equation}
Next we consider the set $\mathcal{E}=\{\boldsymbol{\psi}=\mathbf{s}-\hat{\mathbf{s}}:\mathbf{s},\hat{\mathbf{s}}\in\mathcal{S}\}$,
which consists of all possible difference vectors. Define the parameter
$N$ as the maximum number of leading zeros among these vectors, i.e.,
\begin{equation}
N=\underset{\boldsymbol{\psi}\in\mathcal{E}}{\max}\left\{ \mathfrak{F}\left(\boldsymbol{\psi}\right)\right\} ,\label{eq:Parameter N}
\end{equation}
and the corresponding set $\mathcal{E}_{N}\subset\mathcal{E}$ as
\begin{equation}
\mathcal{E}_{N}=\left\{ \boldsymbol{\psi}\in\mathcal{E}:\mathfrak{F}\left(\boldsymbol{\psi}\right)=N\right\} .\label{eq:set E_N}
\end{equation}
Note that $N$ can equivalently be defined as the maximal number of
leading zeros among all columns of the \ac{SL} matrix $\mathbf{A}$,
i.e., $N=\underset{l\in\left[L\right]}{\max}\left\{ \mathfrak{F}\left(\mathbf{a}_{l}\right)\right\} $,
where $\mathbf{a}_{l}$ denotes the $l$-th column of $\mathbf{A}$.
To see this, first we note that the codebook design condition (\ref{eq:new condition})
implies that for any $\mathbf{s},\hat{\mathbf{s}}\in\mathcal{S}$
with $\mathbf{s}=\sum_{l=1}^{L}\mathbf{s}_{l}$ and $\hat{\mathbf{s}}=\sum_{l=1}^{L}\hat{\mathbf{s}}_{l}$
and $\mathbf{s}\ne\hat{\mathbf{s}}$, for every $m$ we have
\[
s_{m}-\hat{s}_{m}=0\quad\mbox{only if}\quad\mathbf{s}_{l}=\hat{\mathbf{s}}_{l}\quad\forall l\in\mathcal{D}_{m}.
\]
Therefore, (\ref{eq:Function F}) can be restated as {\small{}
\begin{align*}
 & \mathfrak{F}\left(\boldsymbol{\psi}\right)=\mathfrak{F}\left(\mathbf{s}-\hat{\mathbf{s}}\right)=\\
 & \{c-1\ \mathrm{if}\ \mathbf{s}_{l}=\hat{\mathbf{s}}_{l}\ \forall l\in\mathcal{D}_{m},\;m<c\ \mbox{and }\exists\;l\in\mathcal{D}_{c}\;\mbox{such that}\;\mathbf{s}_{l}\neq\hat{\mathbf{s}}_{l}.
\end{align*}
}Next, observe that the maximum in (\ref{eq:Parameter N}) is attained
by pairs of vectors $(\mathbf{s},\hat{\mathbf{s}})$ that satisfy
$\mathbf{s}_{l}=\hat{\mathbf{s}}_{l}$ for all $l$ such that $\mathfrak{F}\left(\mathbf{a}_{l}\right)<N$.
It follows that $\max_{l\in\left[L\right]}\mathfrak{F}(\mathbf{a}_{l})=\max_{\boldsymbol{\psi}\in\mathcal{E}}\mathfrak{F}(\boldsymbol{\psi})=N$.
As an illustrative example, note that for the \ac{SL} matrix given
by (\ref{eq:SL matrix A}), we have $N=2$ since the sixth column
of $\mathbf{A}$ has 2 leading zeros and this is the maximum number
of leading zeros over all columns of $\mathbf{A}$.

Next we provide a high-\ac{SNR} approximation to the \ac{AWEP},
which is based on the fact that at high \ac{SNR}, the \ac{AWEP}
in (\ref{eq:complete AWEP}) is dominated by terms corresponding to
vectors in $\mathcal{E}_{N}$.
\begin{thm}
\label{thm:theorem 1}A high-\ac{SNR} approximation to the \ac{AWEP}
for the SL-MIMO system is given by 
\begin{equation}
\widetilde{\mathrm{AWEP}}\thickapprox\frac{1}{2^{\eta}}\sum_{\mathcal{E}_{N}}\left(\tilde{I}_{1}\left(\boldsymbol{\psi}\right)+\tilde{I}_{2}\left(\boldsymbol{\psi}\right)\right),\label{eq:theorem 1}
\end{equation}
where $\tilde{I}_{1}\left(\boldsymbol{\psi}\right)$ is given by (\ref{eq:asym I1}),{\footnotesize{}}
\begin{figure*}[tbh]
\begin{align}
\tilde{I}_{1}\left(\boldsymbol{\psi}\right)\triangleq & \frac{1}{12C_{N_{t},N_{r}}}\sum_{p\in\mathcal{P}_{N}}\frac{\alpha_{p}\left(1\right)_{\beta_{p,1}}\left(1\right)_{\beta_{p,N+1}}}{2^{\beta_{p,3}+\dots+\beta_{p,n}+n-2}\times\left(1+\frac{1}{2}\right)^{\beta_{p,4}+\dots+\beta_{p,n}+n-3}\times\dots\times\left(1+\frac{1}{N-1}\right)^{\beta_{p,N+1}+\dots+\beta_{p,n}+n-N}}\nonumber \\
 & \times\frac{1}{\left(1+\frac{1+a_{N+1}}{N}\right)^{\beta_{p,N+1}+\dots+\beta_{p,n}+n-N}}\nonumber \\
 & \times\frac{1}{\left(1+\frac{1+a_{N+2}}{N+1+a_{N+1}}\right)^{\beta_{p,N+3}+\dots+\beta_{p,n}+n-N-2}\dots\left(1+\frac{1+a_{n-1}}{n-2+a_{N+1}+\dots+a_{n-2}}\right)^{\beta_{p,n}+1}}\nonumber \\
 & \times\left(\sum_{i_{1}=0}^{\beta_{p,1}}\frac{\left(i_{1}+1\right)_{\beta_{p,2}}}{2^{\beta_{p,2}+i_{1}+1}}.\sum_{i_{2}=0}^{\beta_{p,2}+i_{1}}\frac{\left(i_{2}+1\right)_{\beta_{p,3}}}{\left(1+\frac{1}{2}\right)^{\beta_{p,3}+i_{2}+1}}\dots\sum_{i_{N-1}=0}^{\beta_{p,N-1}+i_{N-2}}\frac{\left(i_{N-1}+1\right)_{\beta_{p,N}}}{\left(1+\frac{1}{N-1}\right)^{\beta_{p,N}+i_{N-1}+1}}\right)\nonumber \\
 & \times\left(\sum_{i_{N+1}=0}^{\beta_{p,N+1}}\frac{\left(i_{N+1}+1\right)_{\beta_{p,N+2}}}{\left(1+\frac{1+a_{N+2}}{N+1+a_{N+1}}\right)^{\beta_{p,N+2}+i_{N+1}+1}}\dots\sum_{i_{n-1}=0}^{\beta_{p,n-1}+i_{n-2}}\frac{\left(i_{n-1}+1\right)_{\beta_{p,n}}}{\left(1+\frac{1+a_{n}}{n-1+a_{N+1}+\dots+a_{n-1}}\right)^{\beta_{p,n}+i_{n-1}+1}}\right).\label{eq:asym I1}\\
\hline \nonumber 
\end{align}
\end{figure*}
 and an expression for $\tilde{I}_{2}\left(\boldsymbol{\psi}\right)$
is obtained by replacing the factor $\frac{1}{12}$ by $\frac{1}{4}$
and the coefficients $a_{m}$ by $b_{m}$, for all $m$, in (\ref{eq:asym I1}).
Note that the set $\mathcal{P}_{N}$ in (\ref{eq:asym I1}) is defined
by 
\begin{equation}
\mathcal{P}_{N}=\left\{ p:B_{p}^{(N)}=B_{\min}^{(N)}\right\} ,\label{eq:The set P}
\end{equation}
where 
\[
B_{\min}^{(N)}=\min_{p\in\left\{ 1,2,\dots,P\right\} }\left\{ B_{p}^{(N)}\right\} ,
\]
and for each $p\in\left\{ 1,2,\ldots,P\right\} $ we have 
\[
B_{p}^{(N)}=\sum_{j=N+1}^{n}\beta_{p,j}.
\]
\end{thm}
\begin{IEEEproof}
See Appendix \ref{sec:Proof-of-Theorem}.
\end{IEEEproof}
From (\ref{eq:asym I1}) we can now analyze the diversity gain of
the system, which is given in the following Proposition.\setcounter{thm}{0}
\begin{prop}
\label{prop:The-diversity-order}The diversity gain of the SL-MIMO
system is given by
\begin{align}
G_{d} & =\left(N_{r}-N\right)\left(N_{t}-N\right).\label{eq:Diversity order}
\end{align}
\end{prop}
\begin{IEEEproof}
Recalling that the values $\left\{ a_{n}\right\} $ are proportional
to the SNR, it can be noticed that the term in the second line of
(\ref{eq:asym I1}) is the only one that does not tend to a constant
value at high \ac{SNR}, i.e., we have 
\[
\left(1+\frac{1+a_{N+1}}{N}\right)^{-x}\rightarrow C\cdot\mathrm{SNR}^{-x}
\]
as $\mathrm{SNR}\rightarrow\infty$, where $C$ is independent of
the SNR. The exponent $x$ thus gives the diversity order of the system.
Considering the set $\mathcal{P}_{N}$, it can be deduced that the
diversity gain of the system is given by 
\begin{equation}
G_{d}=\underset{p}{\min}\left\{ \sum_{j=N+1}^{n}\beta_{p,j}\right\} +n-N.\label{eq:div}
\end{equation}
From (\ref{eq:PDF}), we find that
\begin{align}
B_{\min}^{(N)} & =\underset{p}{\min}\left\{ \sum_{j=N+1}^{n}\beta_{p,j}\right\} =\sum_{j=0}^{n-N-1}\left(\left|N_{t}-N_{r}\right|+2j\right)\nonumber \\
 & =\left(n-N\right)\left|N_{t}-N_{r}\right|+\left(n-N\right)\left(n-N-1\right).\nonumber \\
\label{eq: diversity}
\end{align}
Substituting (\ref{eq: diversity}) into (\ref{eq:div}) we obtain
(\ref{eq:Diversity order}).
\end{IEEEproof}
Note that the diversity gain of the SL-MIMO system is determined
by the number of transmit and receive antennas, and also by the structure
of the \ac{SL} matrix $\mathbf{A}$. Also, we note that for the special
case of X- and Y-codes, our result for the diversity gain matches
that obtained in \cite{vrigneau2008extension,mohammed2011mimo}.

As an illustrative example, the closed-form expression for the approximate
asymptotic \ac{AWEP} for the example system with the \ac{SL} matrix
in (\ref{eq:SL matrix A}) is shown in (\ref{eq:asymptotic AWEP-A}).{\footnotesize{}}
\begin{figure*}[tbh]
\begin{align}
\widetilde{\mathrm{AWEP}}\approx & \frac{1}{2^{\eta}}\sum_{\mathcal{E}_{N}}\left(\left[\frac{1}{12C_{N_{t},N_{r}}}\sum_{p\in\mathcal{P}_{N}}\frac{\alpha_{p}\left(1\right)_{\beta_{p,1}}\left(1\right)_{\beta_{p,3}}}{2^{\beta_{p,3}+\beta_{p,4}+2}\left(1+\frac{1+a_{3}}{2}\right)^{\beta_{p,3}+\beta_{p,4}+2}}\left(\sum_{i_{1}=0}^{\beta_{p,1}}\frac{\left(i_{1}+1\right)_{\beta_{p,2}}}{2^{\beta_{p,2}+i_{1}+1}}\right)\left(\sum_{i_{3}=0}^{\beta_{p,3}}\frac{\left(i_{3}+1\right)_{\beta_{p,4}}}{\left(1+\frac{1+a_{4}}{3+a_{3}}\right)^{\beta_{p,4}+i_{3}+1}}\right)\right]\right.\nonumber \\
 & \left.+\left[\frac{1}{4C_{N_{t},N_{r}}}\sum_{p\in\mathcal{P}_{N}}\frac{\alpha_{p}\left(1\right)_{\beta_{p,1}}\left(1\right)_{\beta_{p,3}}}{2^{\beta_{p,3}+\beta_{p,4}+2}\left(1+\frac{1+b_{3}}{2}\right)^{\beta_{p,3}+\beta_{p,4}+2}}\left(\sum_{i_{1}=0}^{\beta_{p,1}}\frac{\left(i_{1}+1\right)_{\beta_{p,2}}}{2^{\beta_{p,2}+i_{1}+1}}\right)\left(\sum_{i_{3}=0}^{\beta_{p,3}}\frac{\left(i_{3}+1\right)_{\beta_{p,4}}}{\left(1+\frac{1+b_{4}}{3+b_{3}}\right)^{\beta_{p,4}+i_{3}+1}}\right)\right]\right).\label{eq:asymptotic AWEP-A}\\
\hline \nonumber 
\end{align}
\end{figure*}
 This expression is valid for SL-MIMO systems with $n=4$ and $N=2$.
For the \ac{MIMO} system with $N_{t}=N_{r}=4$, the diversity gain
is therefore $G_{d}=4$. It is worth noting that in this case, $p\left(\boldsymbol{\lambda}\right)$
has 201 terms, i.e. $P=201$; however, the asymptotic \ac{AWEP} only
involves 9 of these terms, i.e. $\left|\mathcal{P}_{N}\right|=9$,
with parameters shown in Table~\ref{tab:parameters}.
\begin{table}[tbh]
\caption{Parameters occurring in the expression (\ref{eq:asymptotic AWEP-A})
for the approximate asymptotic \ac{AWEP} of an SL-MIMO system with
$N_{t}=4$, $N_{r}=4$ and $N=2$. Only the terms for $p\in\mathcal{P}_{N}$
are involved (here $|\mathcal{P}_{N}|=9$).\label{tab:parameters}}

\centering{}{\scriptsize{}}%
\begin{tabular}{|>{\centering}p{1cm}|>{\centering}p{1cm}|>{\centering}p{1cm}|>{\centering}p{1cm}|>{\centering}p{1cm}|}
\hline 
\multicolumn{5}{|c|}{{\scriptsize{}$C_{N_{t},N_{r}}=144,\,B_{\min}^{(2)}=\beta_{p,3}+\beta_{p,4}=2$}}\tabularnewline
\hline 
\hline 
{\scriptsize{}$\alpha_{p}$} & {\scriptsize{}$\beta_{p,1}$} & {\scriptsize{}$\beta_{p,2}$} & {\scriptsize{}$\beta_{p,3}$} & {\scriptsize{}$\beta_{p,4}$}\tabularnewline
\hline 
{\scriptsize{}1} & {\scriptsize{}6} & {\scriptsize{}4} & {\scriptsize{}2} & {\scriptsize{}0}\tabularnewline
\hline 
{\scriptsize{}-2} & {\scriptsize{}6} & {\scriptsize{}4} & {\scriptsize{}1} & {\scriptsize{}1}\tabularnewline
\hline 
{\scriptsize{}1} & {\scriptsize{}6} & {\scriptsize{}4} & {\scriptsize{}0} & {\scriptsize{}2}\tabularnewline
\hline 
{\scriptsize{}-2} & {\scriptsize{}5} & {\scriptsize{}5} & {\scriptsize{}2} & {\scriptsize{}0}\tabularnewline
\hline 
{\scriptsize{}4} & {\scriptsize{}5} & {\scriptsize{}5} & {\scriptsize{}1} & {\scriptsize{}1}\tabularnewline
\hline 
{\scriptsize{}-2} & {\scriptsize{}5} & {\scriptsize{}5} & {\scriptsize{}0} & {\scriptsize{}2}\tabularnewline
\hline 
{\scriptsize{}1} & {\scriptsize{}4} & {\scriptsize{}6} & {\scriptsize{}2} & {\scriptsize{}0}\tabularnewline
\hline 
{\scriptsize{}-2} & {\scriptsize{}4} & {\scriptsize{}6} & {\scriptsize{}1} & {\scriptsize{}1}\tabularnewline
\hline 
{\scriptsize{}1} & {\scriptsize{}4} & {\scriptsize{}6} & {\scriptsize{}0} & {\scriptsize{}2}\tabularnewline
\hline 
\end{tabular}{\scriptsize\par}
\end{table}

\section{Codebook Design\label{sec:Constellation-Design}}

In the proposed SL-MIMO system, the codebook $\mathcal{S}_{l}$ for
each layer $l$ is sparsely spread over a subset of the eigen-channels.
Therefore, designing appropriate codebooks $\mathcal{S}_{l}$, as
well as a good sparse layering matrix $\mathbf{A}$, can significantly
improve the error rate performance of the system. In the following,
we focus on codebook optimization with a view to minimizing the average
word error probability of the system.

It is well-known \cite{wang2003simple} that the average error probability
at high \ac{SNR} can be approximated as
\begin{equation}
\mathrm{AWEP}\thickapprox\left(G_{c}\bar{\gamma}\right)^{-G_{d}},\label{eq:ABEP Gc Gd}
\end{equation}
where $\bar{\gamma}$ is the average \ac{SNR}, and $G_{c}$ and $G_{d}$
are the \emph{coding gain} and \emph{diversity gain}, respectively.
As shown in (\ref{eq:Diversity order}), $G_{d}$ depends on the \ac{MIMO}
configuration and the \ac{SL} matrix $\mathbf{A}$, and from (\ref{eq:theorem 1}),
it can be inferred that $G_{c}$ depends on the set $\mathcal{E}_{N}$.
It is well-known that for an $N_{t}\times N_{r}$ \ac{MIMO} system,
the maximum achievable diversity gain is $N_{t}N_{r}$. In the SL-MIMO
system, this maximum can be attained when all layers share the strongest
eigen-channel, i.e., $N=0$. In other words, in a fixed-rate system,
there is a trade-off between diversity gain and coding gain, as sharing
the strongest eigen-channel among all layers results in smaller Euclidean
distances and thus a lower coding gain. Therefore, the desired design
goal can be defined by the optimization problem 
\begin{equation}
\left(\mathbf{A}^{\ast},\{\bar{\mathcal{S}}_{l}^{\ast}\}\right)=\arg\min_{\mathbf{A},\{\bar{\mathcal{S}}_{l}\}}\mathrm{AWEP}_{\mathrm{ub}}\left(\mathbf{A},\{\bar{\mathcal{S}}_{l}\}\right),\label{eq:Optimization problem}
\end{equation}
where $\left(\mathbf{A}^{\ast},\{\bar{\mathcal{S}}_{l}^{\ast}\}\right)$
denotes the jointly optimum values of the \ac{SL} matrix $\mathbf{A}$
and the \emph{shortened codebooks} (i.e., multidimensional constellations)
$\{\bar{\mathcal{S}}_{l}\}$, where $\bar{\mathcal{S}}_{l}$ is the
$n_{l}$-dimensional constellation obtained when we restrict $\mathcal{S}_{l}$
to the index set $\mathcal{N}_{l}$ (i.e., excluding dimensions with
zero values from these vectors), and $\mathrm{AWEP}_{\mathrm{ub}}$
denotes the upper bound on the \ac{AWEP} given by (\ref{eq:complete AWEP}).
The joint optimization in (\ref{eq:Optimization problem}) is prohibitively
complex to implement, even for a small constellation size and a very
limited number of layers and antennas. Therefore, we adopt a more
practical sub-optimal approach which is explained in the following
text. First, we design (or select) an \ac{SL} matrix $\mathbf{A}^{+}$
with the desired number of layers $L$, and then design the optimum
$M$-ary multidimensional constellations $\{\bar{\mathcal{S}}_{l}\}$
for this designed $\mathbf{A}^{+}$. As stated earlier, the complexity
of the \ac{MP} detector is determined by the structure of the \ac{SL}
matrix. Hence, the \ac{SL} matrix can be designed to target a high
diversity gain, while ensuring a good sparse matrix structure for
\ac{MP} decoding.

\begin{figure}[t]
\begin{centering}
\includegraphics[scale=0.15]{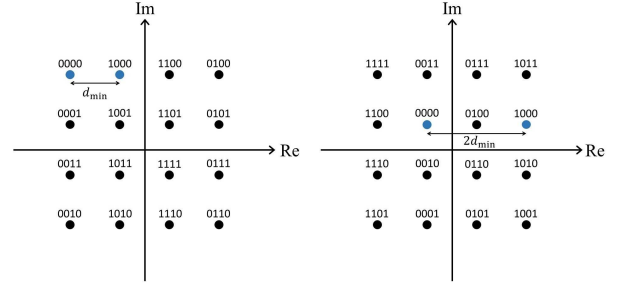}
\par\end{centering}
\caption{An example of a 2-dimensional constellation which uses 16-QAM as the
base constellation $\mathcal{M}$ for the first dimension and a permuted
version of $\mathcal{M}$ for the second dimension. Here the permutation
is performed for the real and imaginary dimensions independently.
Neighboring symbols in the first dimension (eigen-channel) are separated
in the other dimension (eigen-channel). This will increase the overall
minimum distance.\label{fig:16-point-2-dimensional-constella}}
\end{figure}

Having designed the \ac{SL} matrix $\mathbf{A}^{+}$, the optimization
problem is reduced to
\begin{equation}
\{\bar{\mathcal{S}}_{l}^{\ast}\}=\arg\min_{\{\bar{\mathcal{S}}_{l}\}}\mathrm{AWEP}_{\mathrm{ub}}\left(\mathbf{A}^{+},\{\bar{\mathcal{S}}_{l}\}\right),\label{eq:Sub-optimal problem}
\end{equation}
Next, the problem is to jointly design $L$ multidimensional constellations
each having size $M$. It was shown in \cite{boutros1996good} that
the \emph{minimum Euclidean distance} and the \emph{minimum product distance}
of a multidimensional constellation represent the key factors affecting
the error rate performance in Gaussian and Rayleigh fading channels,
respectively. For this reason, most of the multidimensional constellation
design techniques for \ac{SCMA} systems focus on designing a \emph{mother}
constellation for the system to maximize the minimum Euclidean or
product distance of a \emph{single} user (in the SL-MIMO context,
this would correspond to a single layer). A comprehensive comparison
of different multidimensional constellation design methods is presented
in \cite{vameghestahbanati2019multidimensional}. However, in the
proposed SL-MIMO system, the fading experienced in the eigen-channels
corresponds to the \emph{ordered} eigenvalues; due to this fact,
and considering the role of the difference vectors $\boldsymbol{\psi}$
in (\ref{eq:complete AWEP}) and (\ref{eq:asym I1}), it can be seen
that neither the Euclidean distance nor the product distance is a
key factor in determining the error rate performance of SL-MIMO.
On the other hand, it is clear from (\ref{eq:Q function}) that the
\emph{weighted Euclidean distance}, i.e., $\sum_{m}\lambda_{m}\left|\psi_{m}\right|^{2}$,
has a significant effect on the performance of the SL-MIMO system.
However, maximizing the weighted Euclidean distance of the constellation
for a single layer is not a suitable design strategy, as this would
result in dedicating the total energy of each multidimensional symbol
to the strongest available eigen-channel. Therefore, we first choose
a general one-dimensional\footnote{Here we adopt the convention that 'dimensions' refers to \emph{complex}
dimensions, i.e., each dimension corresponds to a complex symbol.} \emph{base constellation} $\mathcal{M}$ with unit average energy,
e.g., a lattice-based constellation such as PAM or QAM, and repeat
it for the other dimensions (the constellation symbol is also repeated
for these dimensions). In addition, permutation among dimensions can
be used such that each multidimensional constellation point contains
a different symbol in each dimension. This helps to increase the minimum
distance\footnote{A similar method has been used for multidimensional constellation
design in \ac{LDS} and \ac{SCMA} systems. Repeated $M$-QAM is identified
as $M$-\ac{LDS} in \cite{hoshyar2008novel}, while the permuted
version was proposed in \cite{taherzadeh2014scma}.} \cite{taherzadeh2014scma}. The permutation can be done by changing
the arrangement in the sequence of constellation points of the base
constellation. Fig.~\ref{fig:16-point-2-dimensional-constella} shows
an example of a 2-dimensional constellation which uses a 16-QAM as
the base constellation, and a permuted version is used for the other
dimension. In this example, the permutation is performed in the real
and imaginary dimensions independently. It can be seen that how permutation
benefits the system via increasing the minimum Euclidean distance:
neighboring symbols in the first dimension (eigen-channel) are separated
in the other dimension (eigen-channel). Any pair of symbols with a
distance of $d_{\min}$ in one dimension has a distance of at least
$2d_{\min}$ in the other dimension.

After choosing a base constellation, we design the \emph{layer operators}
that are applied to the multidimensional constellation in order to
improve the error rate performance by separating the interfering symbols.
Hence, the constellation for a layer $l$ is given by
\[
\bar{\mathcal{S}}_{l}=\left\{ \boldsymbol{\Delta}_{l}\mathbf{m}_{i},i=1,2,\dots,M\right\} ,
\]
where $\boldsymbol{\Delta}_{l}\in\mathbb{C}^{n_{l}\times n_{l}}$
is the $l$-th \emph{layer operator}, and $\mathbf{m}_{i}\in\mathbb{C}^{n_{l}\times1}$
is an $n_{l}$-dimensional symbol whose elements are in the set of
the base constellation $\mathcal{M}$ (note that in repetition mode
the vector $\mathbf{m}_{i}$ is uniquely determined by its first element).

Layer operators are unitary matrices that rotate and scale the complex
symbol to provide unique symbols for each layer. From (\ref{eq:complete AWEP})
it can be seen that $\mathrm{AWEP_{ub}}$ consists of terms of the
form 
\begin{equation}
\frac{\left(k-1+a_{1}+\dots+a_{k-1}\right)^{i_{k-1}}}{\left(k+a_{1}+\dots+a_{k}\right)^{\beta_{p,k}+i_{k-1}+1}}.\label{eq:ratio terms}
\end{equation}
In our scheme, we perform scaling in addition to rotation, since the
fading coefficients are in the ordered condition and scaling factors
help to control $\mathrm{AWEP_{ub}}$ by altering the ratios in (\ref{eq:ratio terms}).
Moreover, the resulting power imbalance helps the \ac{MP} detector
to converge faster, as reliable detection of the ``stronger'' symbols
can aid the detector to subsequently resolve the ``weaker'' symbols.
Hence, the layer operator is written as
\begin{equation}
\boldsymbol{\Delta}_{l}=\left[\begin{array}{ccc}
\rho_{1,l}e^{j\phi_{1,l}} & \ldots & 0\\
\vdots & \ddots & \vdots\\
0 & \ldots & \rho_{n_{l},l}e^{j\phi_{n_{l},l}}
\end{array}\right],l=1,\dots,L,\label{eq:Layer Operator}
\end{equation}
such that
\begin{align}
\rho_{1,l}^{2}+\dots+\rho_{n_{l},l}^{2} & =E_{s},\forall l,\label{eq: rho_l subject to}
\end{align}
where $0\leq\phi_{j,l}<\pi$ for all $j$ and $l$, and $\rho_{j,l}\geq0$
for all $j$ and $l$. Joint optimization of all angles and scaling
factors still requires a high-dimensional exhaustive search of prohibitive
complexity; therefore, to further simplify the problem, first we design
sub-optimal angles $\left\{ \phi_{j,l}^{+}\right\} $ in such a way
as to maximize the minimum Euclidean distance between interfering
symbols independently on each eigen-channel. Then, we proceed to optimize
the scaling factors in order to minimize the error rate. This stage
is also performed in multiple steps, proceeding from the layers using
the weakest eigen-channels to the layers using the strongest eigen-channels,
in order to reduce the complexity of the problem. Then, the optimization
problem of each step of this multi-step problem can be defined as
\begin{align}
\underset{\rho_{j,l},\forall l\in\mathcal{L}_{N'},\forall j\in[n_{l}]}{\mathrm{minimize}} & \mathrm{\widetilde{AWEP}}_{N'}\left(\mathbf{A}^{+},\left\{ \bar{\mathcal{S}}_{l}\left(\boldsymbol{\Delta}_{l}(\rho_{j,l},\phi_{j,l}^{+})\right)\right\} \right)\label{eq:Sub-optimal problem-steps}\\
\mbox{subject to\ \ \ \ } & \mbox{(\ref{eq: rho_l subject to})},\nonumber 
\end{align}
where $\mathcal{L}_{N'}=\left\{ l:\mathfrak{F}\left(\mathbf{a}_{l}\right)=N'\right\} $
and $\mathrm{\widetilde{AWEP}}_{N'}$ is the asymptotic value of the
contribution to the \ac{AWEP} from layers having at least $N'$ leading
zeros in their codewords. $\mathrm{\widetilde{AWEP}}_{N'}$ can be
easily derived from (\ref{eq:theorem 1}) by replacing $N$ with $N'$
in $\tilde{I}_{1}\left(\boldsymbol{\psi}\right)$ and $\tilde{I}_{2}\left(\boldsymbol{\psi}\right)$
and implementing summation over the set $\mathcal{E}_{N'}=\left\{ \boldsymbol{\psi}\in\mathcal{E}:\mathfrak{F}\left(\boldsymbol{\psi}\right)\geq N'\right\} $.
We begin by solving (\ref{eq:Sub-optimal problem-steps}) for $N'=N$,
i.e., $\mathrm{\widetilde{AWEP}}_{N}=\mathrm{\widetilde{AWEP}}$,
and find the corresponding scaling factors $\left\{ \rho_{j,l}\right\} $
by performing a grid search. This grid search operates in parallel
for variables $\rho_{j,l}$ for all $l$ that satisfy $\mathfrak{F}(\mathbf{a}_{l})=N'$
and for all $j=\left\{ 1,2,\dots,n_{l}\right\} $. For each such $l$,
there are $n_{l}-1$ independent variables due to (\ref{eq: rho_l subject to}),
and therefore the number of variables to be optimized in parallel
in this step is $\sum_{l\in\mathcal{L}_{N'}}(n_{l}-1)$, which is
usually manageable since the numbers $n_{l}$ are relatively small.
Then, in each subsequent step, we decrement $N'$ and solve (\ref{eq:Sub-optimal problem-steps})
again for the next set of optimal scaling factors $\left\{ \rho_{j,l}\right\} $.

At the final step we have $N^{\prime}=0$, and $\mathrm{\widetilde{AWEP}}_{0}$
is equivalent to $\mathrm{AWEP_{ub}}$ given in (\ref{eq:complete AWEP}).
At this point, direct optimization of the scaling factors of the remaining
layers using (\ref{eq:complete AWEP}) does not represent a practical
approach as the objective function is relatively complex. Therefore,
recalling the expression (\ref{eq:Q function}) for the \ac{PEP},
we maximize the minimum argument of the Q-function in (\ref{eq:Q function})
by implementing a search process over the layers with $\mathfrak{F}(\mathbf{a}_{l})=0$
, where each eigenvalue $\lambda_{m}$ is replaced by its average
value denoted by $\bar{\lambda}_{m}$. This amounts to maximizing
the minimum\emph{ weighted Euclidean distance} between pairs of symbols,
$\sum_{m}\bar{\lambda}_{m}\left|\psi_{m}\right|^{2}$, where $\left\{ \bar{\lambda}_{m}\right\} $
can be derived analytically from the \ac{MGF} calculated in the previous
section.\footnote{Although this optimization procedure can still have quite a high complexity
for large \ac{MIMO} systems and high-order base constellations $\mathcal{M}$,
it needs to be performed only once and can be deployed offline.} 

\subsection*{Design Procedure}

In this subsection, we briefly summarize the above design procedure
in steps for a generic SL-MIMO system. First, an appropriate sparse
\ac{SL} matrix $\mathbf{A}$ is chosen, taking into account the complexity
of the MP detector and the desired number of layers. Then, we choose
a general $M$-ary one-dimensional \emph{base constellation} $\mathcal{M}$
with unit average energy. This base constellation is repeated (or
permuted) over the other dimensions to produce a generic multidimensional
constellation for the SL-MIMO system. After choosing the \ac{SL}
matrix and constellation, the optimization of the layer operators
in (\ref{eq:Layer Operator}) can be performed using the following
steps.

\emph{Step 1:} We find the angles $\left\{ \phi_{j,l}\right\} $ in
such a way as to maximize the minimum Euclidean distance between interfering
symbols independently on each eigen-channel. Consider the $m$-th
row of $\mathbf{A}$, and note that this row has row weight $d_{m}$,
i.e., $d_{m}$ layers share eigen-channel $m$. As was shown in \cite{van2009multiple},
the optimum rotation angles for $M$-PSK constellation are given by
\begin{equation}
\phi_{j,l}=\frac{k}{d_{m}}\left(\frac{2\pi}{M}\right),\ k=0,\dots,d_{m}-1,\label{eq:phi_j,l-1}
\end{equation}
and therefore this set of rotation angles will be assigned to the
set of interfering symbols.

\emph{Step 2 to end:} We solve the optimization problem in (\ref{eq:Sub-optimal problem-steps})
for $N'=N,N-1,\dots,1,0$. In each step, we find the optimal $\rho_{j,l}$
for all $l$ that satisfy $\mathfrak{F}(\mathbf{a}_{l})=N'$ and for
all $j=\left\{ 1,2,\dots,n_{l}\right\} $. Note that in the final
step when $N'=0$ (and also possibly in preceding steps), since the
objective function in (\ref{eq:Sub-optimal problem-steps}) is relatively
complex, we can instead choose to maximize the minimum\emph{ weighted
Euclidean distance} between pairs of symbols, i.e.,
\[
\underset{\rho_{j,l},\forall l\in\mathcal{L}_{N'},\forall j\in[n_{l}]}{\mbox{maximize}}\left\{ \min_{\boldsymbol{\psi}\in\mathcal{E}_{N'}}\left\{ \sum_{m}\bar{\lambda}_{m}\left|\psi_{m}\right|^{2}\right\} \right\} .
\]

To elucidate this procedure, we next briefly describe the steps for
the design example for the case where the \ac{SL} matrix is given
in (\ref{eq:SL matrix A}), and where the base constellation $\mathcal{M}$
is a 4-QAM (QPSK) constellation.

\emph{Step 1:} Consider rows of $\mathbf{A}$, 3 layers share each
of the eigen-channels. Considering (\ref{eq:phi_j,l-1}), the optimum
rotation angles are $\phi_{j,l}^{+}\in\left\{ 0,\frac{\pi}{6},\frac{\pi}{3}\right\} $;
at each eigen-channel, this set of three rotation angles will be assigned
to the set of three interfering symbols.

\emph{Step 2:} Next, the optimization sub-problem is to minimize $\widetilde{\mathrm{AWEP}}$
in (\ref{eq:asymptotic AWEP-A}) by searching over all possible difference
vectors for layer 6 and find the sub-optimal $\rho_{1,6}^{+}$ and
$\rho_{2,6}^{+}$. Via this method, the design of $\boldsymbol{\Delta}_{6}$
(and thus $\mathcal{S}_{6}$) is achieved. This will be used as the
constellation for layer 6 in the next steps.

\emph{Step 3:} In this step, it is first required to find $\widetilde{\mathrm{AWEP}}_{1}$
as the objective function. This can be found in a way similar to that
of the asymptotic \ac{AWEP}. Referring to (\ref{eq:Sub-optimal problem-steps}),
here our optimization is targeting the layers that satisfy $\mathfrak{F\left(\mathbf{a}_{\mathit{l}}\right)}=1$,
i.e., layers 4 and 5. Therefore, in this step we obtain the scaling
parameters jointly for layers 4 and 5.

\emph{Step 4:} Finally, $N^{\prime}=0$, i.e. the objective function
is the upper bound of \ac{AWEP} given in (\ref{eq:complete AWEP}),
and layers 1, 2 and 3 meet the condition $\mathfrak{F}\left(\mathbf{a}_{l}\right)=0$.
However, instead of solving the problem in (\ref{eq:Sub-optimal problem-steps}),
in this step we maximize the minimum \emph{weighted Euclidean distance}
by searching over all possible difference vectors in order to find
the optimal scaling parameters of $\boldsymbol{\Delta}_{1}$, $\boldsymbol{\Delta}_{2}$
and $\boldsymbol{\Delta}_{3}$.

\section{Numerical Results\label{sec:Numerical-Results}}

\begin{figure}[t]
\centering{}\includegraphics[scale=0.5]{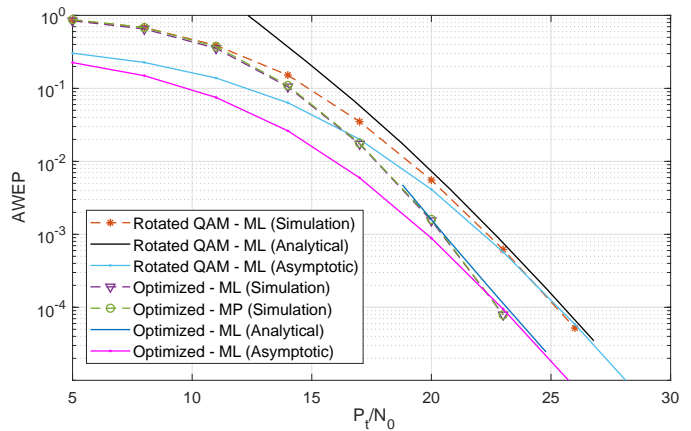}\caption{Illustration of the benefits of the proposed constellation design
method for SL-MIMO. The performance of the SL-MIMO system with the
proposed (optimized) constellations is compared to that of SL-MIMO
with rotated QAM on each layer. Here $N_{t}=N_{r}=4,L=6$ and $M=4$.\label{fig:WEP-performance-full load}}
\end{figure}

\begin{figure}[t]
\begin{centering}
\includegraphics[scale=0.5]{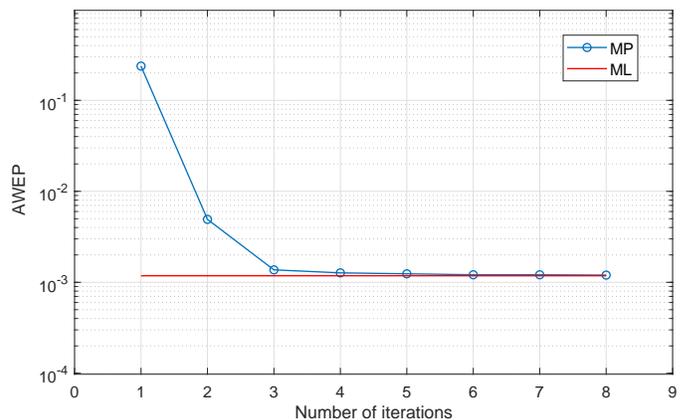}
\par\end{centering}
\caption{Convergence behavior of the \ac{MP} detector for the SL-MIMO system.
Here the \ac{SL} matrix $\mathbf{A}$ is given by (\ref{eq:SL matrix A}),
$N_{t}=N_{r}=4$, $M=4$ and $P_{t}/N_{0}=20\mbox{ dB}$.\label{fig:Convergence-behaviour}}
\end{figure}

In this section, we demonstrate the performance of the proposed SL-MIMO
system design using numerical simulations. Fig.~\ref{fig:WEP-performance-full load}
shows the \ac{AWEP} performance of the SL-MIMO system with $N_{t}=4$,
$N_{r}=4$, $M=4$, $L=6$, and \ac{SL} matrix given by (\ref{eq:SL matrix A}).
The SL-MIMO constellations have been optimized as described in Section~\ref{sec:Constellation-Design}.
For each codebook, a 4-QAM constellation is used as the base constellation
and is only repeated (not permuted) over the second eigen-channel.
The figure provides a comparison with a similar SL-MIMO system where
each layer uses a rotated QAM constellation. It can be observed that
the optimized constellations provide a 2.3~dB improvement at an \ac{AWEP}
of $10^{-4}$. The analytical (equation (\ref{eq:complete AWEP}))
and asymptotic (equation (\ref{eq:asymptotic AWEP-A})) results are
also included in this figure, which verifies the correctness of the
simulation results and also the tightness of the analytical \ac{AWEP}
upper bound. The asymptotic curves in Fig.~\ref{fig:WEP-performance-full load}
verify that the diversity gain of the system is equal to 4, in accordance
with Proposition~\ref{prop:The-diversity-order} (here $N_{t}=N_{r}=4$
and $N=2$). In addition, the results are shown for both the \ac{ML}
and \ac{MP} detection methods. Here 5 iterations were used in the
\ac{MP} detector, as this was found to produce a negligible performance
degradation with respect to \ac{ML} detection, which is shown in
Fig.~\ref{fig:Convergence-behaviour}. This figure presents the \ac{AWEP}
performance of the \ac{MP} detector versus the number of iterations
of message passing, at an \ac{SNR} of 20~dB. It can be seen that
the \ac{MP} detector converges to an \ac{AWEP} value that is very
close to that of the \ac{ML} detector within a very small number
of iterations; hence, from this point onwards, we only plot the performance
of the \ac{MP} detector.

Next we compare the \ac{AWEP} of the proposed SL-MIMO system (with
optimized constellations) with that of X- and Y-codes \cite{mohammed2011mimo},
which serve as the benchmark schemes for the proposed approach. It
was shown in \cite{mohammed2011mimo} that X- and Y-codes perform
better than other competing schemes in this context such as the E-$d_{\min}$
scheme proposed in \cite{vrigneau2008extension} (for more details,
we refer the readers to \cite{mohammed2011mimo} and references therein).
A key advantage of our proposed SL-MIMO scheme is that the system
can be \emph{overloaded}; however, in order to have a fair comparison,
we choose $L=n$, i.e., the number of layers is equal to the number
of available eigen-channels, similar to the case of X- and Y-codes.

\begin{figure}[t]
\begin{centering}
\includegraphics[scale=0.45]{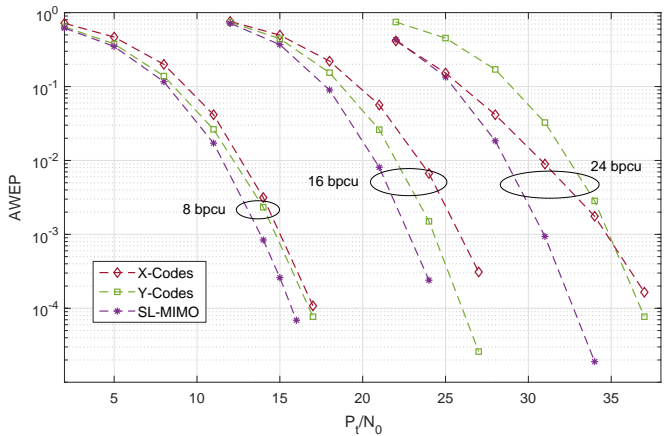}
\par\end{centering}
\caption{Comparison of the \ac{AWEP} performance of the proposed SL-MIMO
system with that of X- and Y-codes for $M=4,16$ and $64$ and $N_{t}=N_{r}=4$.\label{fig:WEP-performance-comparison 4by4}}
\end{figure}

In Fig.~\ref{fig:WEP-performance-comparison 4by4}, we compare the
\ac{AWEP} performance of the proposed SL-MIMO system with $L=4$
with that of X- and Y-codes for $M=4,16\mbox{ and }64$. We use 
\[
\mathbf{A}_{1}=\left[\begin{array}{cccc}
1 & 1 & 1 & 0\\
1 & 0 & 0 & 1\\
0 & 1 & 0 & 1\\
0 & 0 & 1 & 0
\end{array}\right]
\]
 as the \ac{SL} matrix for $M=4$ and $M=16$, where 4-QAM and 16-QAM
are used as the base constellations, respectively, and we use the
matrix 
\[
\mathbf{A}_{2}=\left[\begin{array}{cccc}
1 & 1 & 0 & 0\\
0 & 0 & 1 & 1\\
1 & 0 & 0 & 1\\
0 & 1 & 1 & 0
\end{array}\right]
\]
for the SL-MIMO system with $M=64$, where 8-PAM is used as the base
constellation for the real and imaginary parts of each data symbol.
For each SL-MIMO system, the constellations are optimized as described
in Section~\ref{sec:Constellation-Design}. The results show that
the performance of the SL-MIMO system is superior to that of X- and
Y-codes in all scenarios, and this performance improvement increases
dramatically at higher data rates. This is mainly due to the fact
that in SL-MIMO, a layer interferes with different layers through
its eigen-channels, and the resulting \emph{interference diversity}
benefits the system. The gain over X-codes is approximately 1.2~dB,
2.5~dB and 4.9~dB at 8~bpcu, 16~bpcu and 24~bpcu systems, respectively,
at an \ac{AWEP} of $10^{-4}$. The proposed system also achieves
1~dB, 1.2~dB and 3.2~dB performance improvement over Y-codes at
8~bpcu, 16~bpcu and 24~bpcu, respectively. As noted in \cite{mohammed2011mimo},
we can observe that for low data rates, Y-codes exhibit a performance
improvement over X-codes at all \ac{SNR}, since they perform power
and rate allocation within a pair of layers and eigen-channels; at
higher data rates, however, the performance diminishes significantly,
as instead of using $n$ layers each having $M$ constellation points,
the transmitter for Y-codes creates $\frac{n}{2}$ layers each having
$M^{2}$ constellation points. In fact, an $M^{2}$-QAM symbol is
transmitted by the stronger eigen-channel on each pair and a 2-bit
coded version of that symbol is transmitted through the weaker eigen-channel,
resulting in a large number of difference vectors with small Euclidean
distances, and therefore poor performance at low and medium \acp{SNR}.
In addition, the performance of the X-codes becomes worse at higher
data rates, as the Euclidean distances between interfering QAM symbols
are more sensitive to rotation angles at higher data rates. This results
in Euclidean distances which become close to zero in the medium \ac{SNR}
range, which in turn causes a decrease in the diversity gain.

It is worth mentioning that the complexity of the \ac{MP} detector
is $\mathcal{O}\left(M^{3}\right)$ in the case when $\mathbf{A}_{1}$
is used. However, the complexity of the receiver and optimization
process is reduced by using the matrix $\mathbf{A}_{2}$ for the SL-MIMO
system with $M=64$. In addition to this, instead of optimizing the
rotation angles in \emph{Step~1}, we set $\phi_{j,l}=0$ in order
to obtain $\emph{real-valued}$ $\boldsymbol{\Delta}_{l}$. This further
reduces the complexity of the receiver by separating the real and
imaginary parts of the data symbols and using two $\sqrt{M}$-PAM
constellations together with application of two independent detectors
at the receiver on the real and imaginary parts of the received signal.
Hence, the complexity of the receiver in this case is $\mathcal{O}\left(M\right)$,
similar to the case of X-codes.

\begin{figure}[t]
\begin{centering}
\includegraphics[scale=0.45]{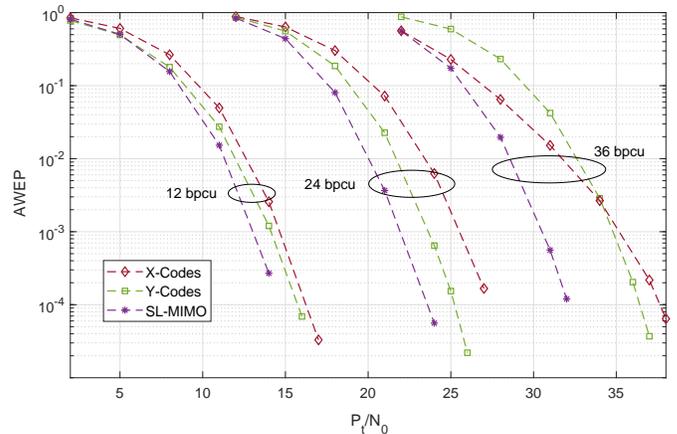}
\par\end{centering}
\caption{Comparison of the \ac{AWEP} performance of the proposed SL-MIMO
system with that of X- and Y-codes for $M=4,16$ and $64$ and $N_{t}=N_{r}=6$.\label{fig:WEP-performance-comparison 6by6}}
\end{figure}

\begin{figure*}[tbh]
\begin{centering}
\includegraphics[scale=0.4]{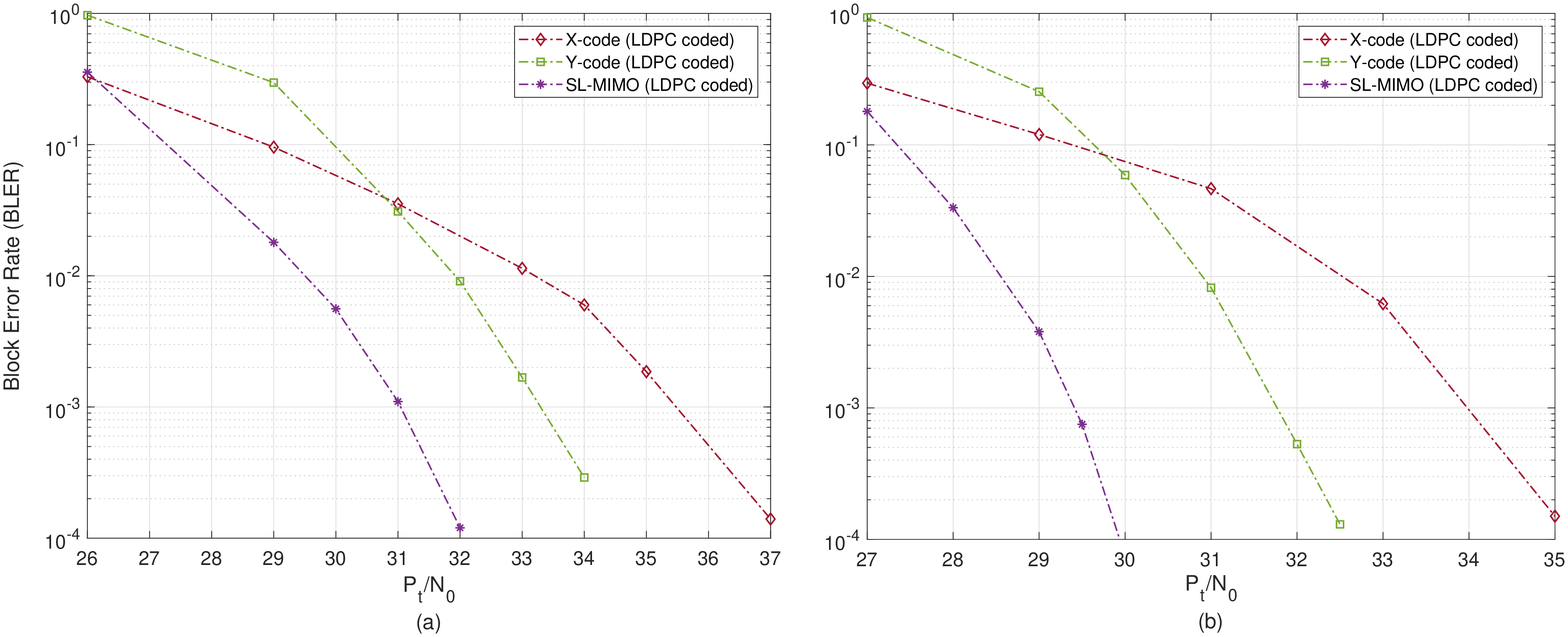}
\par\end{centering}
\caption{Comparison of the average BLER performance of the proposed SL-MIMO
system with that of X- and Y-codes for $M=64$, where an LDPC code
is used with rate 0.47. (a) $N_{t}=N_{r}=4$, (b) $N_{t}=N_{r}=6$.\label{fig:BlEP-LDPC_coded}}
\end{figure*}

Fig.~\ref{fig:WEP-performance-comparison 6by6} compares the performance
of the SL-MIMO system with X- and Y-codes in a $6\times6$ \ac{MIMO}
system, i.e., $N_{t}=N_{r}=6$. Similar to the $4\times4$ system,
we use two different \ac{SL} matrices,{\small{}
\begin{equation}
\mathbf{A}_{3}=\left[\begin{array}{cccccc}
1 & 1 & 1 & 0 & 0 & 0\\
0 & 0 & 0 & 1 & 1 & 1\\
0 & 0 & 1 & 0 & 0 & 1\\
0 & 1 & 0 & 0 & 1 & 0\\
1 & 0 & 0 & 1 & 0 & 0\\
0 & 0 & 0 & 0 & 0 & 0
\end{array}\right],\mathbf{A}_{4}=\left[\begin{array}{cccccc}
1 & 1 & 0 & 0 & 0 & 0\\
0 & 0 & 1 & 1 & 0 & 0\\
0 & 0 & 0 & 0 & 1 & 1\\
0 & 0 & 0 & 1 & 1 & 0\\
0 & 1 & 1 & 0 & 0 & 0\\
1 & 0 & 0 & 0 & 0 & 1
\end{array}\right],\label{eq:connctivity matrices A3 and A4}
\end{equation}
}$\mathbf{A}_{3}$ for $M=4$ and 16, and $\mathbf{A}_{4}$ for $M=64$.
Similar to the $4\times4$ system, we use 4- and 16-QAM constellations
for the cases where $M=4$ and $M=16$, respectively, and for the
system with $M=64$, 8-PAM is used as the base constellation per real
dimension. As expected, the SL-MIMO system achieves even higher gains
in a larger \ac{MIMO} system. The constellation-optimized SL-MIMO
designs provide 1.6~dB, 3.7~dB and 5.6~dB improvement over X-codes
and 1.2~dB, 1.7~dB and 4.7~dB gain over Y-codes at 12~bpcu, 24~bpcu
and 36~bpcu systems, respectively, at an \ac{AWEP} of $10^{-4}$.
Recalling (\ref{eq:connctivity matrices A3 and A4}), the complexity
of the \ac{MP} detector in the $6\times6$ system is similar to that
of the $4\times4$ system. It can be seen that in $\mathbf{A}_{3}$,
3 layers are interfering at eigen-channels 1 and 2, therefore $d_{\max}=$3
and thus, the complexity of the receiver is $\mathcal{O}\left(M^{3}\right)$.
However, in the matrix $\mathbf{A}_{4}$, each eigen-channel is shared
between 2 layers, and since $\phi_{j,l}=0$, and consequently the
real and imaginary parts of data symbols are separated, thus the complexity
of the receiver is $\mathcal{O}\left(M\right)$.

In Fig.~\ref{fig:BlEP-LDPC_coded}, we compare the error rate performance
of SL-MIMO systems and X- and Y-codes in the presence of channel
coding. Here we apply the LDPC code with code rate 0.47 that has been
implemented in the 5G New Radio (NR) standard \cite{etsiTS138212},
and the constellation size is $M=64$. The results show that the proposed
SL-MIMO scheme remains dominant in error rate performance over X-
and Y-codes when channel coding is implemented in the considered system.
The gain over X-codes is 5~dB and 5.5~dB, and the gain over Y-codes
is 2.5~dB and 2.8~dB in $4\times4$ and $6\times6$ MIMO systems,
respectively, at an average block error rate (BLER) of $10^{-4}$.
We also observe that the resulting enhancement is increasing with
an increase in the SNR.

\section{Conclusion\label{sec:Conclusion}}

In this paper, we proposed the SL-MIMO transmission technique, which
improves the error rate performance of \ac{MIMO} systems and thereby
provides excellent performance in both diversity order and coding
gain. In SL-MIMO, code-domain non-orthogonal transmission and \ac{SVD}
precoding are combined to allow re-use of eigen-channels. At the receiver,
a low-complexity \ac{MP} detector can be applied which has performance
close to that of \ac{ML} detection. We evaluated the error rate performance
of the SL-MIMO system by calculating the joint \ac{MGF} of the \emph{ordered}
eigenvalues. A closed-form upper bound on the \ac{AWEP} and asymptotic
\ac{AWEP} were derived and the diversity gain of the system was analytically
obtained. Based on these analytical results, an optimization procedure
was introduced to design sparse layering codebooks to minimize the
error rate of the SL-MIMO system. The error rate performance of the
SL-MIMO system has been compared with that of X- and Y-codes by numerical
simulations. The results have shown that the performance of the SL-MIMO
system outperforms other competing schemes in all of the considered
scenarios. This enhancement becomes more prominent at higher data
rates and in larger \ac{MIMO} systems. An interesting direction of
future research along this line is that of extension of the SL-MIMO
scheme to a \emph{multiuser} uplink communication scenario, where
users play the role of layers, and in particular the development of
suitable low-complexity message-passing detection algorithms for the
multiuser SL-MIMO setting.

\appendices{}

\section{Proof of Theorem \ref{thm:theorem 1}\label{sec:Proof-of-Theorem}}

To prove (\ref{eq:asym I1}), we rewrite (\ref{eq:int I1}) in the
equivalent form of (\ref{eq:newformat I1}),{\footnotesize{}}
\begin{figure*}[tbh]
\begin{align}
I_{1}\left(\boldsymbol{\psi}\right)= & \frac{1}{12C_{N_{t},N_{r}}}\sum_{p=1}^{P}\frac{\alpha_{p}\left(1\right)_{\beta_{p,1}}}{\left(1+a_{1}\right)^{\beta_{p,1}+1}}\times\frac{1}{\left(1+a_{1}\right)^{\beta_{p,2}+1}\left(2+a_{1}+a_{2}\right)^{\beta_{p,3}+1}\dots\left(n-1+a_{1}+\dots+a_{n-1}\right)^{\beta_{p,n}+1}}\nonumber \\
 & \times\sum_{i_{1}=0}^{\beta_{p,1}}\frac{\left(i_{1}+1\right)_{\beta_{p,2}}}{\left(1+\frac{1+a_{2}}{1+a_{1}}\right)^{\beta_{p,2}+i_{1}+1}}\sum_{i_{2}=0}^{\beta_{p,2}+i_{1}}\frac{\left(i_{2}+1\right)_{\beta_{p,3}}}{\left(1+\frac{1+a_{3}}{2+a_{1}+a_{2}}\right)^{\beta_{p,3}+i_{2}+1}}\dots\sum_{i_{n-1}=0}^{\beta_{p,n-1}+i_{n-2}}\frac{\left(i_{n-1}+1\right)_{\beta_{p,n}}}{\left(1+\frac{1+a_{n}}{n-1+a_{1}+\dots+a_{n-1}}\right)^{\beta_{p,n}+i_{n-1}+1}}.\nonumber \\
\label{eq:newformat I1}\\
\hline \nonumber 
\end{align}
\end{figure*}
 which (with some minor manipulation) can be re-expressed as (\ref{eq:newformat I1-1}).
\begin{figure*}[tbh]
\begin{align}
I_{1}\left(\boldsymbol{\psi}\right)= & \frac{1}{12C_{N_{t},N_{r}}}\sum_{p=1}^{P}\frac{\alpha_{p}\left(1\right)_{\beta_{p,1}}}{\left(1+a_{1}\right)^{\beta_{p,1}+\dots+\beta_{p,n}+n}\left(1+\frac{1+a_{2}}{1+a_{1}}\right)^{\beta_{p,3}+\dots+\beta_{p,n}+n-2}\dots\left(1+\frac{1+a_{n-1}}{n-2+a_{1}+\dots+a_{n-2}}\right)^{\beta_{p,n}+1}}\nonumber \\
 & \times\sum_{i_{1}=0}^{\beta_{p,1}}\frac{\left(i_{1}+1\right)_{\beta_{p,2}}}{\left(1+\frac{1+a_{2}}{1+a_{1}}\right)^{\beta_{p,2}+i_{1}+1}}\sum_{i_{2}=0}^{\beta_{p,2}+i_{1}}\frac{\left(i_{2}+1\right)_{\beta_{p,3}}}{\left(1+\frac{1+a_{3}}{2+a_{1}+a_{2}}\right)^{\beta_{p,3}+i_{2}+1}}\dots\sum_{i_{n-1}=0}^{\beta_{p,n-1}+i_{n-2}}\frac{\left(i_{n-1}+1\right)_{\beta_{p,n}}}{\left(1+\frac{1+a_{n}}{n-1+a_{1}+\dots+a_{n-1}}\right)^{\beta_{p,n}+i_{n-1}+1}}.\nonumber \\
\label{eq:newformat I1-1}\\
\hline \nonumber 
\end{align}
\end{figure*}

It can be seen in (\ref{eq:newformat I1-1}) that there are terms
of the form 
\[
\left(1+\frac{1+a_{k}}{k-1+a_{1}+\dots+a_{k-1}}\right)^{-c_{p,k}}
\]
in the first line with $k=2,\dots,n-1$ and $c_{p,k}=\beta_{p,k+1}+\dots+\beta_{p,n}+n-k$,
and in the nested sum in the second line with $k=2,\dots,n$ and $c_{p,k}=\beta_{p,k}+i_{k-1}+1$.
For a difference vector $\boldsymbol{\psi}$ which has no nonzero
entries, each of these terms tends to a constant at high SNR, and
only the term $\left(1+a_{1}\right)^{-(\beta_{p,1}+\dots+\beta_{p,n}+n)}$
in line 1 varies with \ac{SNR}. However, for a difference vector
$\boldsymbol{\psi}$ with $\mathfrak{F}\left(\boldsymbol{\psi}\right)=m>0$
(i.e., $a_{1}=\dots=a_{m}=0$), only the term with $k=m+1$, i.e.
$\left(1+\frac{1+a_{m+1}}{m}\right)^{-c_{p}}$, where $c_{p}=\beta_{p,m+1}+\dots+\beta_{p,n}+n-m+i_{m}$,
varies at high \ac{SNR}. In addition, for each $p$, the minimum
value of $c_{p}$ among all vectors $\boldsymbol{\psi}\in\mathcal{E}$
corresponds to the maximum value of $m$, which is equal to $N$.
Hence, it can be deduced that vectors $\boldsymbol{\psi}\in\mathcal{E}_{N}$
provide terms that dominate the high-\ac{SNR} value of the \ac{AWEP}.
As a consequence, to obtain the approximate \ac{AWEP}, we set $\psi_{1}=\dots=\psi_{N}=0$,
or equivalently $a_{1}=\dots=a_{N}=0$, in equation (\ref{eq:newformat I1-1});
after this we obtain (\ref{eq: I1 a_i=00003D0}).
\begin{figure*}[tbh]
\begin{align}
 & I_{1}\left(\boldsymbol{\psi}|\psi_{1}=\dots=\psi_{N}=0\right)=\nonumber \\
 & \frac{1}{12C_{N_{t},N_{r}}}\sum_{p=1}^{P}\frac{\alpha_{p}\left(1\right)_{\beta_{p,1}}}{2^{\beta_{p,3}+\dots+\beta_{p,n}+n-2}\times\left(1+\frac{1}{2}\right)^{\beta_{p,4}+\dots+\beta_{p,n}+n-3}\times\dots\times\left(1+\frac{1}{N-1}\right)^{\beta_{p,N+1}+\dots+\beta_{p,n}+n-N}}\nonumber \\
 & \times\frac{1}{\left(1+\frac{1+a_{N+1}}{N}\right)^{\beta_{p,N+2}+\dots+\beta_{p,n}+n-N-1}}\nonumber \\
 & \times\frac{1}{\left(1+\frac{1+a_{N+2}}{N+1+a_{N+1}}\right)^{\beta_{p,N+3}+\dots+\beta_{p,n}+n-N-2}\dots\left(1+\frac{1+a_{n-1}}{n-2+a_{N+1}+\dots+a_{n-2}}\right)^{\beta_{p,n}+1}}\nonumber \\
 & \times\sum_{i_{1}=0}^{\beta_{p,1}}\frac{\left(i_{1}+1\right)_{\beta_{p,2}}}{2^{\beta_{p,2}+i_{1}+1}}\sum_{i_{2}=0}^{\beta_{p,2}+i_{1}}\frac{\left(i_{2}+1\right)_{\beta_{p,3}}}{\left(1+\frac{1}{2}\right)^{\beta_{p,3}+i_{2}+1}}\dots\sum_{i_{N-1}=0}^{\beta_{p,N-1}+i_{N-2}}\frac{\left(i_{N-1}+1\right)_{\beta_{p,N}}}{\left(1+\frac{1}{N-1}\right)^{\beta_{p,N}+i_{N-1}+1}}\nonumber \\
 & \times\sum_{i_{N}=0}^{\beta_{p,N}+i_{N-1}}\frac{\left(i_{N}+1\right)_{\beta_{p,N+1}}}{\left(1+\frac{1+a_{N+1}}{N}\right)^{\beta_{p,N+1}+i_{N}+1}}\nonumber \\
 & \times\sum_{i_{N+1}=0}^{\beta_{p,N+1}+i_{N}}\frac{\left(i_{N+1}+1\right)_{\beta_{p,N+2}}}{\left(1+\frac{1+a_{N+2}}{N+1+a_{N+1}}\right)^{\beta_{p,N+2}+i_{N+1}+1}}\dots\sum_{i_{n-1}=0}^{\beta_{p,n-1}+i_{n-2}}\frac{\left(i_{n-1}+1\right)_{\beta_{p,n}}}{\left(1+\frac{1+a_{n}}{n-1+a_{N+1}+\dots+a_{n-1}}\right)^{\beta_{p,n}+i_{n-1}+1}}.\label{eq: I1 a_i=00003D0}\\
\hline \nonumber 
\end{align}
\end{figure*}

From (\ref{eq: I1 a_i=00003D0}), we can observe that the terms in
line 2 and 5 are constants, and the terms in line 4 and the nested
sum in line 7 also tend to constants which depend on the value of
the vector $\boldsymbol{\psi}$. In addition, the terms of the sum
in line 6 are of the form 
\[
\left(\omega_{1}+\omega_{2}\mathrm{SNR}\right)^{-\delta},
\]
 where $\delta=\beta_{p,N+1}+i_{N}+1$. At high-\ac{SNR} values,
the sum can be approximated by the term with the largest exponent
(smallest $\delta$), which corresponds to the term with $i_{N}=0$.
Hence, it can be concluded that
\begin{align}
\sum_{i_{N}=0}^{\beta_{p,N}+i_{N-1}}\frac{\left(i_{N}+1\right)_{\beta_{p,N+1}}}{\left(1+\frac{1+a_{N+1}}{N}\right)^{\beta_{p,N+1}+i_{N}+1}}\nonumber \\
\rightarrow\frac{\left(1\right)_{\beta_{p,N+1}}}{\left(1+\frac{1+a_{N+1}}{N}\right)^{\beta_{p,N+1}+1}}.\label{eq:approximate sum}
\end{align}
Substituting (\ref{eq:approximate sum}) into (\ref{eq: I1 a_i=00003D0}),
we obtain (\ref{eq:asym I1}). The restriction of values of $p$ to
the set $\mathcal{P}_{N}$, defined in (\ref{eq:The set P}), is due
to the fact that the high-\ac{SNR} approximation of the sum over
$p$ is dominated by the terms with the largest exponent of \ac{SNR}
(line 2 in (\ref{eq:asym I1})), which are the terms with the minimum
value of $\left(\beta_{p,N+1}+\dots+\beta_{p,n}\right)$.

\bibliographystyle{ieeepes}
\bibliography{ref}

\end{document}